\documentclass{article}
\usepackage{graphicx}
\usepackage{amsmath}

\input{tcilatex}

\begin{document}

\title{\textbf{Dual Quantum Mechanics}}
\author{W. Chagas-Filho \\
Physics Department, Federal University of Sergipe, Brazil}
\maketitle

\begin{abstract}
We point out a possible complementation of the basic equations of quantum
mechanics in the presence of gravity.This complementation is suggested by
the well-known fact that quantum mechanics can be equivalently formulated in
the position or in the momentum representation. As a way to support this
complementation, starting from the action that describes conformal gravity
in the world-line formalism, we show that there are duality transformations
that relate the dynamics in the presence of position dependent vector and
tensor fields to the dynamics in the presence of momentum dependent vector
and tensor fields.
\end{abstract}

\section{Introduction}

\noindent \noindent The wave-particle duality of matter and energy at the
quantum level is one of the most fundamental aspects of physics. The far
reaching theoretical implications of the existence of the wave-particle
duality are not completely understood until now. The best known implication
of this duality is the fact that quantum mechanics can be equivalently
formulated in the position representation or in the momentum representation.
While the position representation emphasizes the particle aspect by assuming
a defined position, the momentum representation is related to the wave
aspect because the magnitude $p$ of the momentum of a particle is directly
related to the wave length $\lambda $ of the associated wave by the de
Broglie relation $p=\frac{h}{\lambda }$, where $h$ is Planck%
\'{}%
s constant.

Some years ago, it was discovered [1] that this duality of the descriptions
in terms of position and momentum in quantum mechanics has a symmetric
version as a local $Sp(2,R)$ symmetry of a classical action describing
conformal gravity on the world-line. Local $Sp(2,R)$ symmetry treats
position and momentum as indistinguishable variables and, for this, it
requires conformal gravity in a space-time with an extra space-like
dimension and an extra time-like dimension [1]-[18]. Extra space-like
dimensions had previously been found in string theory, but this was the
first time that an extra time-like dimension was explicitly found. For this
reason, this area of research is sometimes referred to as two-time (2T)
physics.

The important aspect of 2T physics is that we can always use the local
invariance of the action to eliminate the extra dimensions and work with the
emergent gauge-fixed theory, containing only the physical degrees of
freedom, and therefore avoiding the ghost problem in the quantized theory.
Using this approach, it was demonstrated [14], [19] that the Standard Model
of Particles and Forces and General Relativity as we know them are only
holographic shadows of a more symmetric theory with one extra space-like
dimension and one extra time-like dimension. For this reason, it is
important to try to understand other aspects of the gravitational physics
with two time-like dimensions. Following this point of view, in this paper
we present a number of $(d+2)$-dimensional constrained Hamiltonian
formalisms starting from the first order conformal gravity action in the
world-line formalism. The new aspect of these Hamiltonian formalisms is that
they are connected by duality transformations that interchange position and
momentum. Dualities of this kind play a significant role in M-theory, and
because of the wave-particle duality one may expect this same kind of
duality to appear in quantum gravity. We use the dual classical Hamiltonian
formalisms we present in this paper as a basis to suggest a complementation
of the basic equations of quantum mechanics in the presence of gravity.

2T physics is usually considered as an approach that provides a new
perspective for understanding one-time dynamics (1T physics) from a higher
dimensional, more unified point of view. In this paper we are not interested
in this 2T to 1T holographic property [7] of 2T physics. What we have in
mind in this paper is the fact that all the fundamental interactions are
described by gauge theories, and the fact that all gauge theories can be
described as constrained Hamiltonian systems with first class constraints
[20], [21].

In papers [5], [7], [19] it was required that the local invariance of the
conformal gravity action must be the $Sp(2,R)$ invariance even when
space-time fields are present. To satisfy this requirement, in these papers,
the space-time gravitational and vector fields must satisfy certain
conditions that can not be derived from the action. This uncomfortable
situation is avoided later by performing a transition to the field theory
formalism [14], [19], where an action is introduced from which the
conditions leading to local $Sp(2,R)$ invariance can be derived. However,
finding a natural way of introducing space-time tensor and vector fields in
conformal gravity in the world-line formalism remained an open problem until
now. In this paper we present an initial attempt to solve this problem.
Although our attempt does not require local $Sp(2,R)$ symmetry, being
instead based on the presence of another local symmetry of the action, it
brings with it interesting new insights into the internal structure of
quantum mechanics.

The $Sp(2,R)$ symmetry is the local symmetry of conformal gravity,
discovered for the world-line action in the absence of interactions in the $%
(d+2)$-dimensional flat space-time. However, $Sp(2,R)$ is no longer the
local symmetry when fermions are introduced in the formalism. It is
substituted [2], [5] by local $OSp(n\mid 2)$, which contains $Sp(2,R)$ and
reduces to it when the fermions are removed from the formalism. This result
may be viewed as suggesting a possible solution to the problem of
introducing space-time fields in the world-line formalism. Instead of
requiring local $Sp(2,R)$ symmetry of the action with space-time fields as a
starting point, we can try to find another local symmetry that can be used
to eliminate the same number of unphysical degrees of freedom that are
eliminated using the local $Sp(2,R)$ symmetry, thus also avoiding the ghost
problem in the quantized theory. A consistency condition is that this new
local symmetry must reproduce the local $Sp(2,R)$ symmetry when the
space-time fields are absent. We will describe in this paper how, starting
from the first order conformal gravity action in the world-line formalism,
we can construct a natural constrained Hamiltonian formalism, with first
class constraints, containing space-time tensor and vector fields. In this
Hamiltonian formalism, the first class constraints generate local symmetries
that reproduce the $Sp(2,R)$ local symmetry when the space-time fields are
absent. The formalism displays duality transformations that change the
dynamics with position dependent tensor and vector fields into the dynamics
with momentum dependent tensor and vector fields, and vice versa.

In the usual 1T physics, position dependent tensor fields play an important
role in the most general position space formulation of quantum mechanics in
the presence of gravity [22]. In this general formulation, these tensor
fields appear in the spectral decomposition of the unity, define the correct
integration measure for the inner product and are present in the most
general expression of the position matrix elements for self adjoint momentum
operators in position space [22] 
\begin{equation*}
\langle x\mid \hat{p}_{\alpha }\mid x%
{\acute{}}%
\rangle =\frac{i\hbar }{g^{\frac{1}{4}}(x)}\frac{\partial }{\partial
x^{\alpha }}[\frac{1}{g^{\frac{1}{4}}(x)}\delta ^{n}(x-x%
{\acute{}}%
)]
\end{equation*}
\begin{equation}
+\frac{1}{\sqrt{g(x)}}A_{\alpha }(x)\delta ^{n}(x-x%
{\acute{}}%
)  \tag{1.1}
\end{equation}
where $g(x)=\det g_{\alpha \beta }(x)$ and $\alpha ,\beta =1,...,n$. Since
quantum mechanics can be equivalently formulated in the position or in the
momentum representation, the appearance of a momentum dependent tensor field
in conformal gravity can be interpreted as an indication that the momentum
space versions of quantum mechanical equations such as (1.1) and others are
still lacking. The construction and the justification of these momentum
space equations are the motivations for this paper.

As can be seen in (1.1), the other central object in the general position
space formulation of quantum mechanics described in [22] is the vector field 
$A_{\alpha }(x)$. It has a vanishing strength tensor, 
\begin{equation}
F_{\alpha \beta }=\frac{\partial A_{\beta }}{\partial x^{\alpha }}-\frac{%
\partial A_{\alpha }}{\partial x^{\beta }}=0  \tag{1.2}
\end{equation}
and because of this condition it defines a section of a flat U(1) bundle
over the position space. The vector field is present only if the position
space has a non-trivial topology. In position spaces with trivial topology $%
A_{\alpha }(x)$ can always be gauged away [22].

In this paper we describe how one can extend to momentum space the general
position space formulation of quantum mechanics described in [22]. The paper
is organized as follows. In section two we describe how duality
transformations relate position dependent and momentum dependent tensor
fields in $d$-dimensional relativistic massless particle theory in a
constrained Hamiltonian framework.

In section three we review the global and local symmetries of the conformal
gravity action.

Section four presents the basic equations of a formulation of quantum
mechanics that completely incorporates the wave-particle duality in the
presence of gravity. This is done by introducing the corresponding momentum
space expressions of the position space expressions obtained in [22].

In section five we present the developments that suggest our construction in
section four. The starting point is the action describing conformal gravity
on the world-line. In section 5.1 we present an action in a position
dependent tensor background and compute the conserved Hamiltonian Noether
charge corresponding to the local invariance of the action. We find that the
conserved Noether charge and the equations of motion reproduce the conserved
charge and the equations of motion of conformal gravity in a transition to
flat space. In section 5.2 we study the same situation for an action in a
momentum dependent tensor background and find exactly the same behavior of
the conserved Noether charge and of the equations of motion. In addition, a
duality transformation changes the equations of motion in the momentum
dependent background into the equations of motion in the position dependent
background obtained in section 5.1. In section 5.3 we consider the case of
flat space with a position dependent vector field. In section 5.4 we
consider the case of flat space with a momentum dependent vector field.
Again we discover that the conserved charges and the equations of motion
reproduce those of conformal gravity when the vector field vanishes, and
that the equations of motion in the presence of the vector field are turned
into one another by a duality transformation. Section 5.5 considers the case
when both the tensor and vector fields are present, and we are lead to
identical conclusions about the conserved charges and equations of motion.
All the actions we compute in this section describe the correct number of
physical degrees of freedom, thus avoiding the ghost problem in the
quantized theory. Concluding remarks appear in section six.

\section{Massless Relativistic Particles}

In this section we consider massless relativistic particle theory. We
describe how a duality transformation relates the local symmetry and the
equations of motion in a position dependent background to the local symmetry
and equations of motion in a momentum dependent background.

A massless scalar relativistic particle in a $d$-dimensional Minkowski
space-time with signature $(d-1,1)$ is described by the action 
\begin{equation}
S=\frac{1}{2}\int d\tau \lambda ^{-1}\dot{x}^{2}  \tag{2.1}
\end{equation}
where $\lambda (\tau )$ is an auxiliary variable, $x^{\mu }=x^{\mu }(\tau )$%
, $\dot{x}^{2}=\dot{x}^{\mu }\dot{x}^{\nu }\eta _{\mu \nu }$ and $\eta _{\mu
\nu }$, with $\mu ,\nu =0,1,...,d-1$, is the flat Minkowski metric. A dot
denotes derivatives with respect to the arbitrary parameter $\tau $. Action
(2.1) is invariant under the local infinitesimal reparametrizations 
\begin{equation}
\delta x_{\mu }=\alpha (\tau )\dot{x}_{\mu }\text{ \ \ \ \ \ }\delta \lambda
=\frac{d}{d\tau }[\alpha (\tau )\lambda ]  \tag{2.2}
\end{equation}
where $\alpha (\tau )$ is an arbitrary function. Due to the presence of this
local invariance action (2.1) can be looked at as describing gravity on the
world-line. It is well known that action (2.1) also has a global $d$
dimensional conformal invariance. Global conformal invariance in $d$
dimensions is isomorphic to global Lorentz invariance in $d+2$ dimensions
[23[, [24]. Therefore, there is a $d+2$ dimensional Lorentz invariant
extension of action (2.1). This higher dimensional action is the subject of
the next section.

In the transition to the Hamiltonian formalism action (2.1) gives the
canonical momenta 
\begin{equation}
p_{\lambda }=0  \tag{2.3}
\end{equation}
\begin{equation}
p_{\mu }=\frac{\dot{x}_{\mu }}{\lambda }  \tag{2.4}
\end{equation}
and the Hamiltonian 
\begin{equation}
H=\frac{1}{2}\lambda p^{2}  \tag{2.5}
\end{equation}
Equation (2.3) is a primary constraint [20]. Introducing the Lagrange
multiplier $\xi (\tau )$ for this constraint we can write the total
Hamiltonian [20] 
\begin{equation}
H_{T}=H+\xi p_{\lambda }=\frac{1}{2}\lambda p^{2}+\xi p_{\lambda }  \tag{2.6}
\end{equation}
Introducing the Poisson bracket $\{\lambda ,p_{\lambda }\}=1$ and requiring
the dynamical stability [20] of constraint (2.3) 
\begin{equation}
\dot{p}_{\lambda }=\{p_{\lambda },H_{T}\}=0  \tag{2.7}
\end{equation}
we obtain the secondary constraint 
\begin{equation}
\phi =\frac{1}{2}p^{2}\approx 0  \tag{2.8}
\end{equation}
Constraint (2.8) needs not be incorporated into the formalism because it
already appears in the Hamiltonian. Requiring its dynamical stability 
\begin{equation}
\dot{\phi}=\{\phi ,H_{T}\}=0  \tag{2.9}
\end{equation}
we find that it is automatically satisfied. Constraints (2.3) and (2.8) have
vanishing Poisson bracket and are therefore first class constraints [20].
Constraint (2.3) generates translations in the arbitrary variable $\lambda
(\tau )$ and can be dropped from the formalism. The notation $\approx 0$
means that $\phi $ \textsl{weakly vanishes} [20], [21].

Action (2.1) can be rewritten in Hamiltonian form as 
\begin{equation}
S=\int d\tau (\dot{x}.p-\frac{1}{2}\lambda p^{2})  \tag{2.10}
\end{equation}
By introducing the Poisson brackets 
\begin{equation}
\{x_{\mu },x_{\nu }\}=0\text{ \ \ \ \ }\{p_{\mu },p_{\nu }\}=0\text{ \ \ \ \ 
}\{x_{\mu },p_{\nu }\}=\eta _{\mu \nu }  \tag{2.11}
\end{equation}
we can check that the first class constraint (2.8) generates the local
transformations with arbitrary parameter $\epsilon (\tau )$ 
\begin{equation}
\delta x_{\mu }=\epsilon (\tau )\{x_{\mu },\phi \}=\epsilon p_{\mu } 
\tag{2.12a}
\end{equation}
\begin{equation}
\delta p_{\mu }=\epsilon (\tau )\{p_{\mu },\phi \}=0  \tag{2.12b}
\end{equation}
under which action (2.10) transforms as 
\begin{equation}
\delta S=\int d\tau \lbrack \frac{d}{d\tau }(\epsilon \phi )+\dot{\epsilon}%
\phi -\phi \delta \lambda ]  \tag{2.12c}
\end{equation}
If we choose $\delta \lambda =\dot{\epsilon}$ the variation (2.12c) becomes 
\begin{equation}
\delta S=\int d\tau \frac{d}{d\tau }(\epsilon \phi )  \tag{2.12d}
\end{equation}
In this case the quantity 
\begin{equation}
Q=\epsilon \phi =\frac{1}{2}\epsilon p^{2}  \tag{2.13}
\end{equation}
can be interpreted as the conserved Hamiltonian Noether charge, or as the
generator of the local symmetry transformations (2.12), depending on wether
the equations of motion are satisfied or not [25]. Using the local
invariance generated by the charge (2.13), it is possible [21] to eliminate
the time-like degree of freedom by a gauge-fixing. This leaves us with $d-1$
canonical pairs describing the physical degrees of freedom. In this case
there will be no negative norm states (ghosts) in the quantized theory. As
we will see in the following, all the actions we discuss in this paper
describe this same number of physical canonical pairs.

Hamiltonian (2.5) generates the equations of motion. 
\begin{equation}
\dot{x}_{\mu }=\{x_{\mu },H\}=\lambda p_{\mu }  \tag{2.14a}
\end{equation}
\begin{equation}
\dot{p}_{\mu }=\{p_{\mu },H\}=0  \tag{2.14b}
\end{equation}

\subsection{Position dependent tensor fields}

Action (2.10) has an extension in curved space given by [19] 
\begin{equation}
S=\int d\tau \lbrack \dot{x}^{\mu }p_{\mu }-\frac{1}{2}\lambda g_{\mu \nu
}(x)p^{\mu }p^{\nu }]  \tag{2.15}
\end{equation}
where the Hamiltonian is 
\begin{equation}
H=\frac{1}{2}\lambda g_{\mu \nu }(x)p^{\mu }p^{\nu }  \tag{2.16}
\end{equation}
The position dependent tensor field $g_{\mu \nu }(x)$ is defined over phase
space. The equation of motion for $\lambda (\tau )$ gives the constraint 
\begin{equation}
\phi =\frac{1}{2}g_{\mu \nu }(x)p^{\mu }p^{\nu }\approx 0  \tag{2.17}
\end{equation}
Requiring the dynamical stability [20] condition $\dot{\phi}=\{\phi
,H_{T}\}=0$ for constraint (2.17), we find that it is automatically
satisfied. This implies that constraint (2.17) is a first class constraint
[20].

Constraint (2.17) generates the local transformations 
\begin{equation}
\delta x_{\mu }=\epsilon (\tau )\{x_{\mu },\phi \}=\epsilon g_{\mu \nu
}(x)p^{\nu }  \tag{2.18a}
\end{equation}
\begin{equation}
\delta p_{\mu }=\epsilon (\tau )\{p_{\mu },\phi \}=-\frac{1}{2}\epsilon 
\frac{\partial g_{\alpha \beta }(x)}{\partial x^{\mu }}p^{\alpha }p^{\beta }
\tag{2.18b}
\end{equation}
under which 
\begin{equation}
\delta S=\int d\tau \lbrack \frac{d}{d\tau }(\epsilon \phi )+\dot{\epsilon}%
\phi -\phi \delta \lambda ]  \tag{2.18c}
\end{equation}
If we then choose $\delta \lambda =\dot{\epsilon}$ we get 
\begin{equation}
\delta S=\int d\tau \frac{d}{d\tau }(\epsilon \phi )  \tag{2.18d}
\end{equation}
This shows that the conserved Noether charge corresponding to the local
symmetry transformations (2.18) in the background $g_{\mu \nu }(x)$ is the
quantity 
\begin{equation}
Q=\epsilon \phi =\frac{1}{2}\epsilon g_{\mu \nu }(x)p^{\mu }p^{\nu } 
\tag{2.19}
\end{equation}
Using the local symmetry generated by the conserved charge (2.19) it is
possible [21] to eliminate the time-like degrees of freedom. This leaves
only $d-1$ physical canonical pairs. In this case the physical components of
the tensor field will depend only on the physical components of the position
variable. There will be no ghosts in the quantized theory.

Hamiltonian (2.16) generates the equations of motion 
\begin{equation}
\dot{x}_{\mu }=\{x_{\mu },H\}=\lambda g_{\mu \nu }(x)p^{\nu }  \tag{2.20a}
\end{equation}
\begin{equation}
\dot{p}_{\mu }=\{p_{\mu },H\}=-\frac{1}{2}\lambda \frac{\partial g_{\alpha
\beta }(x)}{\partial x^{\mu }}p^{\alpha }p^{\beta }  \tag{2.20b}
\end{equation}
The equations of motion (2.20) reproduce the equations of motion (2.14) when 
$g_{\mu \nu }(x)=\eta _{\mu \nu }$.

\subsection{Momentum dependent tensor fields}

Now we apply to action (2.15) the duality transformation 
\begin{equation}
x_{\mu }(\tau )\rightarrow p_{\mu }(\tau )\text{ \ \ \ \ \ \ \ }p_{\mu
}(\tau )\rightarrow -x_{\mu }(\tau )  \tag{2.21}
\end{equation}
This duality transformation leaves invariant the definition $\{A,B\}=\frac{%
\partial A}{\partial x^{\mu }}\frac{\partial B}{\partial p_{\mu }}-\frac{%
\partial A}{\partial p_{\mu }}\frac{\partial B}{\partial x^{\mu }}$ of the
classical Poisson bracket between two functions $A$ and $B$ of the canonical
variables, and also the fundamental commutator $[x_{\mu },p_{\nu }]=x_{\mu
}p_{\nu }-p_{\nu }x_{\mu }=ih\eta _{\mu \nu }$ of quantum mechanics. We
obtain the action 
\begin{equation}
S=\int d\tau \lbrack -x^{\mu }\dot{p}_{\mu }-\frac{1}{2}\lambda g_{\mu \nu
}(p)x^{\mu }x^{\nu }]  \tag{2.22}
\end{equation}
where the Hamiltonian is 
\begin{equation}
H=\frac{1}{2}\lambda g_{\mu \nu }(p)x^{\mu }x^{\nu }  \tag{2.23}
\end{equation}
and $g_{\mu \nu }(p)$ describes a momentum dependent tensor field defined
over phase space. The equation of motion for the variable $\lambda (\tau )$
gives the constraint 
\begin{equation}
\phi =\frac{1}{2}g_{\mu \nu }(p)x^{\mu }x^{\nu }\approx 0  \tag{2.24}
\end{equation}
Requiring the dynamical stability [20] of constraint (2.24), we find that it
is automatically satisfied and that constraint (2.24) is a first class
constraint [20]. It generates the local transformations 
\begin{equation}
\delta x_{\mu }=\epsilon (\tau )\{x_{\mu },\phi \}=\frac{1}{2}\epsilon \frac{%
\partial g_{\alpha \beta }(p)}{\partial p^{\mu }}x^{\alpha }x^{\beta } 
\tag{2.25a}
\end{equation}
\begin{equation}
\delta p_{\mu }=\epsilon (\tau )\{p_{\mu },\phi \}=-\epsilon g_{\mu \alpha
}(p)x^{\alpha }  \tag{2.25b}
\end{equation}
under which 
\begin{equation}
\delta S=\int d\tau \lbrack \frac{d}{d\tau }(\epsilon \phi )+\dot{\epsilon}%
\phi -\phi \delta \lambda ]  \tag{2.25c}
\end{equation}
If we choose $\delta \lambda =\dot{\epsilon}$ the variation (2.25c) becomes 
\begin{equation}
\delta S=\int d\tau \frac{d}{d\tau }(\epsilon \phi )  \tag{2.25d}
\end{equation}
This demonstrates that the conserved Noether charge in the background $%
g_{\mu \nu }(p)$ is the quantity 
\begin{equation}
Q=\epsilon \phi =\frac{1}{2}\epsilon g_{\mu \nu }(p)x^{\mu }x^{\nu } 
\tag{2.26}
\end{equation}
Using the local symmetry generated by the conserved charge (2.26) it is
possible [21] to eliminate the time-like degrees of freedom. This again
leaves only $d-1$ physical canonical pairs. The physical components of the
tensor field will depend only on the physical components of the momentum
variable, and there will be no ghosts in the quantized theory. The $d$%
-dimensional massless particle actions (2.10), (2.15) and (2.22) therefore
describe the dynamics of the same number of physical canonical pairs.

Notice that the local symmetry transformations (2.25) and the corresponding
conserved charge (2.26) in the background $g_{\mu \nu }(p)$ are turned by
the duality transformation (2.21) into the local symmetry transformations
(2.18) and the corresponding conserved charge (2.19) in the background $%
g_{\mu \nu }(x).$ Although the duality transformation (2.21) is not a
symmetry of actions (2.15) and (2.22), it relates the local symmetries of
these two actions.

Hamiltonian (2.23) generates the equations of motion 
\begin{equation}
\dot{x}_{\mu }=\{x_{\mu },H\}=\frac{1}{2}\lambda \frac{\partial g_{\alpha
\beta }(p)}{\partial p^{\mu }}x^{\alpha }x^{\beta }  \tag{2.27a}
\end{equation}
\begin{equation}
\dot{p}_{\mu }=\{p_{\mu },H\}=-\lambda g_{\mu \alpha }(p)x^{\alpha } 
\tag{2.27b}
\end{equation}
Notice that the equations of motion (2.27) are turned by the duality
transformation (2.21) into the equations of motion (2.20). The
transformation (2.21) relates two possible Hamiltonian descriptions of the
dynamics in the presence of tensor backgrounds. It is this dual behavior of
the local symmetries and of the dynamical evolutions in the presence of
space-time fields that we want to reproduce in $d+2$ dimensions. For this
purpose we must first show that our $(d+2)$-dimensional actions have a local
symmetry that can be used to eliminate three unphysical degrees of freedom
from each of the canonical variables, leaving only $d-1$ physical canonical
pairs. Local $Sp(2,R)$ is not an acceptable symmetry in this case because it
treats position and momentum as indistinguishable variables. To search for
another local symmetry, we must first understand how local $Sp(2,R)$ works.
This is the subject of the next section.

\section{Conformal Gravity and 2T Physics}

The construction of 2T physics [1]-[18] is based on the introduction of a
new gauge invariance in phase space by gauging the duality (2.21) for the
quantum commutator $[X_{M},P_{N}]=ih\eta _{MN}$ with $M,N=0,1,...,d+1$. This
procedure leads to a symplectic $Sp(2,R)$ gauge theory. To remove the
distinction between position and momentum we rename them $%
X_{1}^{M}=X^{M}(\tau )$ and $X_{2}^{M}=P^{M}(\tau )$ and define the doublet $%
X_{i}^{M}(\tau )=(X_{1}^{M},X_{2}^{M})$. The local $Sp(2,R)$ symmetry acts
as [1] 
\begin{equation}
\delta X_{i}^{M}(\tau )=\epsilon _{ik}\omega ^{kl}(\tau )X_{l}^{M}(\tau ) 
\tag{3.1}
\end{equation}
$\omega ^{ij}(\tau )$ is a symmetric matrix containing three local
parameters and $\epsilon _{ij}$ is the Levi-Civita symbol that serves to
raise or lower indices. The $Sp(2,R)$ gauge field $A^{ij}$ is symmetric in $%
(i,j)$ and transforms as [1] 
\begin{equation}
\delta A^{ij}=\partial _{\tau }\omega ^{ij}+\omega ^{ik}\epsilon
_{kl}A^{lj}+\omega ^{jk}\epsilon _{kl}A^{il}  \tag{3.2}
\end{equation}
The covariant derivative is [1] 
\begin{equation}
D_{\tau }X_{i}^{M}=\partial _{\tau }X_{i}^{M}-\epsilon _{ik}A^{kl}X_{l}^{M} 
\tag{3.3}
\end{equation}
An action invariant under the local $Sp(2,R)$ symmetry is [1] 
\begin{equation*}
S=\frac{1}{2}\int d\tau (D_{\tau }X_{i}^{M})\epsilon ^{ij}X_{j}^{N}\eta _{MN}
\end{equation*}
\begin{equation*}
=\int d\tau \lbrack \frac{1}{2}(\partial _{\tau
}X_{1}^{M}X_{2}^{N}-X_{1}^{M}\partial _{\tau }X_{2}^{N})-\frac{1}{2}%
A^{ij}X_{i}^{M}X_{j}^{N}]\eta _{MN}
\end{equation*}
\begin{equation}
=\int d\tau \lbrack \frac{1}{2}(\dot{X}^{M}P_{M}-X^{M}\dot{P}_{M})-(\frac{1}{%
2}\lambda _{1}P^{2}+\lambda _{2}X.P+\frac{1}{2}\lambda _{3}X^{2})] 
\tag{3.4a}
\end{equation}
where $A^{11}=\lambda _{3}$, $\ A^{12}=A^{21}=\lambda _{2}$, \ $%
A^{22}=\lambda _{1}$. After an integration by parts this action can be
written as 
\begin{equation}
S=\int d\tau \lbrack \dot{X}^{M}P_{M}-(\frac{1}{2}\lambda _{1}P^{2}+\lambda
_{2}X.P+\frac{1}{2}\lambda _{3}X^{2})]  \tag{3.4b}
\end{equation}
From the form (3.4b) for the action we can identify the Hamiltonian as 
\begin{equation}
H=\frac{1}{2}\lambda _{1}P^{2}+\lambda _{2}X.P+\frac{1}{2}\lambda _{3}X^{2} 
\tag{3.5}
\end{equation}
The first-order Hamiltonian action (3.4b) describes conformal gravity on the
world-line [1], [26], [27]. We can obtain the usual second-order Lagrangian
action for conformal gravity by solving the equation of motion for $P_{M}$
that follows from action (3.4b) and inserting the solution back into it.

The equations of motion for the variables $\lambda _{\alpha }(\tau )$, $%
\alpha =1,2,3$ give the constraints 
\begin{equation}
\phi _{1}=\frac{1}{2}P^{2}\approx 0  \tag{3.6a}
\end{equation}
\begin{equation}
\phi _{2}=X.P\approx 0  \tag{3.6b}
\end{equation}
\begin{equation}
\phi _{3}=\frac{1}{2}X^{2}\approx 0  \tag{3.6c}
\end{equation}
and therefore the $\lambda _{\alpha }$ are arbitrary variables. Constraints
(3.6) can only be simultaneously satisfied in a flat space-time with two
time-like dimensions [1]-[18]. Constraints (3.6) were independently obtained
in [23].

We now introduce the Poisson brackets 
\begin{equation}
\{P_{M},P_{N}\}=0\text{ \ \ \ \ }\{X_{M},X_{N}\}=0\text{ \ \ \ \ }%
\{X_{M},P_{N}\}=\eta _{MN}  \tag{3.7}
\end{equation}
and require the dynamical stability [20] of constraints (3.6) 
\begin{equation}
\dot{\phi}_{\alpha }=\{\phi _{\alpha },H\}=\lambda _{\beta }\{\phi _{\alpha
},\phi _{\beta }\}=0  \tag{3.8}
\end{equation}
We then obtain the bracket relations 
\begin{equation}
\{\phi _{1},\phi _{2}\}=-2\phi _{1}  \tag{3.9a}
\end{equation}
\begin{equation}
\{\phi _{1},\phi _{3}\}=-\phi _{2}  \tag{3.9b}
\end{equation}
\begin{equation}
\{\phi _{2},\phi _{3}\}=-2\phi _{3}  \tag{3.9c}
\end{equation}
The bracket relations (3.9) weakly vanish, indicating that the stability
conditions (3.8) are automatically satisfied and that constraints (3.6) are
first class constraints. The algebra (3.9) is the $Sp(2,R)$ gauge algebra.

Action (3.4b) is invariant under global Lorentz $SO(d,2)$ transformations
with generator $L_{MN}=X_{M}P_{N}-X_{N}P_{M}$ 
\begin{equation}
\delta X_{M}=\frac{1}{2}\omega _{RS}\{L_{RS},X_{M}\}=\omega _{MR}X_{R} 
\tag{3.10a}
\end{equation}
\begin{equation}
\delta P_{M}=\frac{1}{2}\omega _{RS}\{L_{RS},P_{M}\}=\omega _{MR}P_{R} 
\tag{3.10b}
\end{equation}
\begin{equation}
\delta \lambda _{\alpha }=0,\text{ \ \ \ \ \ \ }\alpha =1,2,3  \tag{3.10c}
\end{equation}
under which $\delta S=0$. The $L_{MN}$ are gauge invariant because they have
vanishing brackets with constraints (3.6). The remarkable consequence of
this is that the global Lorentz $SO(d,2)$ symmetry survives the gauge-fixing
process and is present in all the gauge-fixed systems. Perhaps the most
striking example of this is the hidden Lorentz $SO(d,2)$ symmetry of the
non-relativistic massive particle (see ref. [4] for details).

The first class constraints (3.6) generate local infinitesimal $Sp(2,R)$
transformations with arbitrary parameters $\epsilon _{\alpha }(\tau )$ 
\begin{equation}
\delta X_{M}=\epsilon _{\alpha }(\tau )\{X_{M},\phi _{\alpha }\}=\epsilon
_{1}P_{M}+\epsilon _{2}X_{M}  \tag{3.11a}
\end{equation}
\begin{equation}
\delta P_{M}=\epsilon _{\alpha }(\tau )\{P_{M},\phi _{\alpha }\}=-\epsilon
_{2}P_{M}-\epsilon _{3}X_{M}  \tag{3.11b}
\end{equation}
under which 
\begin{equation*}
\delta S=\int d\tau \lbrack (\epsilon _{2}\lambda _{1}-\epsilon _{1}\lambda
_{2})2\phi _{1}+(\epsilon _{3}\lambda _{1}-\epsilon _{1}\lambda _{3})\phi
_{2}
\end{equation*}
\begin{equation}
+(\epsilon _{3}\lambda _{2}-\epsilon _{2}\lambda _{3})2\phi _{3}+\frac{d}{%
d\tau }(\epsilon _{\alpha }\phi _{\alpha })+\dot{\epsilon}_{\alpha }\phi
_{\alpha }-\phi _{\alpha }\delta \lambda _{\alpha }]  \tag{3.11c}
\end{equation}
Using the constraint equations (3.6) we can write 
\begin{equation}
\delta S\approx \int d\tau \lbrack \frac{d}{d\tau }(\epsilon _{\alpha }\phi
_{\alpha })+\dot{\epsilon}_{\alpha }\phi _{\alpha }-\phi _{\alpha }\delta
\lambda _{\alpha }]  \tag{3.11d}
\end{equation}
We can see from (3.11d) that if we now choose $\delta \lambda _{\alpha }=%
\dot{\epsilon}_{\alpha }$ the variation of the action becomes 
\begin{equation}
\delta S\approx \int d\tau \lbrack \frac{d}{d\tau }(\epsilon _{\alpha }\phi
_{\alpha })]  \tag{3.11e}
\end{equation}
Equation (3.11e) indicates that, in conformal gravity in the world-line
formalism, the quantity 
\begin{equation}
Q=\epsilon _{\alpha }\phi _{\alpha }=\frac{1}{2}\epsilon _{1}P^{2}+\epsilon
_{2}X.P+\frac{1}{2}\epsilon _{3}X^{2}  \tag{3.12}
\end{equation}
can be interpreted as the conserved Hamiltonian Noether charge or as the
generator of the local infinitesimal transformations (3.11), depending on
wether the equations of motion are satisfied or not [25]. Using the local
symmetry generated by the conserved charge (3.12) we can eliminate one
space-like degree of freedom and two time-like degrees of freedom of each of
the canonical variables [1]-[19]. In this case we are left with $d-1$
physical canonical pairs, which is the same number of physical canonical
pairs we found for the $d$-dimensional massless particle in the previous
section. There will be no ghosts in the quantized theory.

Action (3.4a) is invariant under the duality transformation 
\begin{equation}
X_{M}(\tau )\rightarrow P_{M}(\tau )\text{ \ \ \ \ \ \ }P_{M}(\tau
)\rightarrow -X_{M}(\tau )  \tag{3.13a}
\end{equation}
\begin{equation}
\lambda _{1}\rightarrow \lambda _{3}\text{ \ \ \ \ }\lambda _{2}\rightarrow
-\lambda _{2}\text{ \ \ \ \ }\lambda _{3}\rightarrow \lambda _{1} 
\tag{3.13b}
\end{equation}
If we perform the duality transformation (3.13) with the $\epsilon _{\alpha
} $ in the place of the $\lambda _{\alpha }$ we find that the local
transformation equations (3.11a), (3.11b) and the conserved charge (3.12)
remain invariant.

Hamiltonian (3.5) generates the equations of motion 
\begin{equation}
\dot{X}_{M}=\{X_{M},H\}=\lambda _{1}P_{M}+\lambda _{2}X_{M}  \tag{3.14a}
\end{equation}
\begin{equation}
\dot{P}_{M}=\{P_{M},H\}=-\lambda _{2}P_{M}-\lambda _{3}X_{M}  \tag{3.14b}
\end{equation}
The equations of motion (3.14) are invariant under the duality
transformation (3.13). Now we clearly see what are the consequences of local 
$Sp(2,R)$ symmetry: it leaves the local transformation equations and the
equations of motion invariant under dualities of the type (3.13).

But this invariance of the equations of motion under the duality
transformation (3.13a) does not exist in quantum mechanics. The quantum
operators $\hat{x}_{\mu }=x_{\mu }$, \ $\hat{p}_{\mu }=ih\frac{\partial }{%
\partial x^{\mu }}$ are turned into the operators $\hat{p}_{\mu }=p_{\mu }$,
\ $\hat{x}_{\mu }=-ih\frac{\partial }{\partial p^{\mu }}$, and the equations
of motion in the position representation are turned into the equations of
motion in the momentum representation under (3.13). For this reason, we will
not require local $Sp(2,R)$ symmetry for our actions in the presence of
vector and tensor fields. For our purposes here we need to find another
local symmetry. This other symmetry must lead to the correct number of
physical canonical pairs and must reproduce the local $Sp(2,R)$ symmetry in
the absence of space-time fields. This will be the subject of section five.
Let us now consider how we can introduce a complementation of the basic
equations of quantum mechanics in the presence of gravity.

\section{Dual Quantum Mechanics}

The results in this paper bring with them the possibility of a deeper
insight into the structure of quantum mechanics. The idea is to further
incorporate into quantum mechanics the duality of the descriptions in terms
of position and momentum. To this end we must introduce an additional
assumption between assumptions A1 and A2 of reference [22]. The
complementation of quantum mechanics we present in this section is based on
three assumptions, of which the first and the third ones are identical to A1
and A2 in [22]. Our assumptions are

1) There exists a basis $\mid x\rangle $ of the position space which is
spanned by the eigenvalues of the position operators $\hat{x}^{\alpha }$ $%
(\alpha =1,2,...,n),$ whose domain of eigenvalues coincides with all the
possible values of the coordinates $x^{\alpha }$ parameterizing the position
space $M(x)$, 
\begin{equation*}
\hat{x}^{\alpha }\mid x\rangle =x^{\alpha }\mid x\rangle \text{ \ \ , \ \ }%
\{x^{\alpha }\}\in M(x)
\end{equation*}

2) There exists a basis $\mid p\rangle $ of the momentum space which is
spanned by the eigenvalues of the momentum operators $\hat{p}_{\alpha }$ $%
(\alpha =1,2,...,n),$ whose domain of eigenvalues coincides with all the
possible values of the momenta $p_{\alpha }$ parameterizing the momentum
space $D(p)$, 
\begin{equation*}
\hat{p}_{\alpha }\mid p\rangle =p_{\alpha }\mid p\rangle \text{ \ , \ }%
\{p_{\alpha }\}\in D(p)
\end{equation*}

3) The representation spaces of the algebra are endowed with a Hermitian
positive definite inner product $\langle .\mid .\rangle $ for which the
operators $\hat{x}^{\alpha }$ and $\hat{p}_{\alpha }$ are self-adjoint.

Now consider quantum mechanics in the position representation. The
construction of quantum mechanics describing the diffeomorphic-covariant
representations of the Heisenberg algebra in terms of topological classes of
a flat U(1) bundle over position space has the parameterization [22] of the
inner product $\langle x\mid x^{\prime }\rangle $ 
\begin{equation}
\langle x\mid x^{\prime }\rangle =\frac{1}{\sqrt{g(x)}}\delta
^{n}(x-x^{\prime })  \tag{4.1}
\end{equation}
where $g(x)$ is an arbitrary positive definite function defined over the
position space $M$. For a Riemannian manifold the natural choice [22] for $g$
is the determinant of the metric tensor, $g=\det g_{\alpha \beta }(x)$.
Position dependent metric tensors naturally appear in this formulation of
quantum mechanics.

Equation (4.1) implies the spectral decomposition [22] of the identity
operator in the position eigenbasis $\mid x\rangle $ 
\begin{equation}
1=\int_{M}d^{n}x\sqrt{g(x)}\mid x\rangle \langle x\mid  \tag{4.2}
\end{equation}
which in turn leads to the position space wave function representations $%
\psi (x)=\langle x\mid \psi \rangle $ and $\langle \psi \mid x\rangle
=\langle x\mid \psi \rangle ^{\ast }=\psi ^{\ast }(x)$ of any state $\mid
\psi \rangle $ belonging to the Heisenberg algebra representation space, 
\begin{equation}
\mid \psi \rangle =\int_{M}d^{n}x\sqrt{g(x)}\psi (x)\mid x\rangle  \tag{4.3}
\end{equation}
\begin{equation}
\langle \psi \mid =\int_{M}d^{n}x\sqrt{g(x)}\psi ^{\ast }(x)\langle x\mid 
\tag{4.4}
\end{equation}
The inner product of two states $\mid \psi \rangle $ and $\mid \varphi
\rangle $ is then given in terms of their position space wave functions $%
\psi (x)$ and $\varphi (x)$ as 
\begin{equation}
\langle \psi \mid \varphi \rangle =\int_{M}d^{n}x\sqrt{g(x)}\psi ^{\ast
}(x)\varphi (x)  \tag{4.5}
\end{equation}
The most general position space wave function representations of the
position and momentum operators are [22] 
\begin{equation}
\langle x\mid \hat{x}_{\alpha }\mid \psi \rangle =x_{\alpha }\langle x\mid
\psi \rangle =x_{\alpha }\psi (x)  \tag{4.6a}
\end{equation}
\begin{equation}
\langle x\mid \hat{p}_{\alpha }\mid \psi \rangle =\frac{-i\hbar }{g^{1/4}(x)}%
\left[ \frac{\partial }{\partial x^{\alpha }}+\frac{i}{\hbar }A_{\alpha }(x)%
\right] g^{1/4}(x)\psi (x)  \tag{4.6b}
\end{equation}
The vector field $A_{\alpha }(x)$ is present only in the case of
topologically non-trivial position spaces [22]. It has a vanishing strength
tensor $F_{\alpha \beta }$ as given by (1.2) and is related to arbitrary
local phase transformations of the position eigenvectors 
\begin{equation}
\mid x^{\prime }\rangle =e^{\frac{i}{\hbar }\chi (x)}\mid x\rangle 
\tag{4.7a}
\end{equation}
when 
\begin{equation}
A_{\alpha }^{\prime }(x)=A_{\alpha }(x)+\frac{\partial \chi (x)}{\partial
x^{\alpha }}  \tag{4.7b}
\end{equation}
where $\chi (x)$ is an arbitrary scalar function. From the above equations
we see that position dependent tensors play a central role in this
formulation of quantum mechanics. The other central object in this
formulation is the vector field $A_{\alpha }(x)$ of vanishing strength
tensor.

Now consider quantum mechanics in the momentum representation. The
normalization of the momentum eigenstates is parameterized according to [22] 
\begin{equation}
\langle p\mid p^{\prime }\rangle =\frac{1}{\sqrt{h(p)}}\delta
^{n}(p-p^{\prime })  \tag{4.8}
\end{equation}
where $h(p)$ is an arbitrary positive definite function defined over the
domain $D(p)$ of the momentum eigenvalues. The authors in [22] do not go
beyond this point and do not consider the possible forms of the function $%
h(p)$. However, as a consequence of the wave-particle duality, we may expect
the form $h(p)=\det g_{\alpha \beta }(p)$ to be a possible one.

As a consequence of (4.8) and of our second assumption, we have the spectral
decomposition of the identity operator in the momentum eigenbasis $\mid
p\rangle $ 
\begin{equation}
1=\int_{D(p)}d^{n}p\sqrt{h(p)}\mid p\rangle \langle p\mid  \tag{4.9}
\end{equation}
This leads to the momentum space wave functions $\psi (p)=\langle p\mid \psi
\rangle $ and $\langle \psi \mid p\rangle =\langle p\mid \psi \rangle ^{\ast
}=\psi ^{\ast }(p)$ of any state $\mid \psi $ $\rangle $ belonging to the
Heisenberg algebra representation space 
\begin{equation}
\mid \psi \rangle =\int_{D(p)}d^{n}p\sqrt{h(p)}\psi (p)\mid p\rangle 
\tag{4.10a}
\end{equation}
\begin{equation}
\langle \psi \mid =\int_{D(p)}d^{n}p\sqrt{h(p)}\psi ^{\ast }(p)\langle p\mid
\tag{4.10b}
\end{equation}
The inner product of two states $\mid \psi \rangle $ and $\mid \varphi
\rangle $ is given in terms of their momentum space wave functions $\psi (p)$
and $\varphi (p)$ as 
\begin{equation}
\langle \psi \mid \varphi \rangle =\int_{D(p)}d^{n}p\sqrt{h(p)}\psi ^{\ast
}(p)\varphi (p)  \tag{4.11}
\end{equation}

The most general wave function $\langle x\mid p\rangle $ is given by [22] 
\begin{equation}
\langle x\mid p\rangle =\frac{e^{i\varphi (x_{0},p)}}{(2\pi \hbar )^{\frac{n%
}{2}}}\frac{\Omega \lbrack P(x_{0}\rightarrow x)]}{g^{\frac{1}{4}}(x)h^{%
\frac{1}{4}}(p)}e^{\frac{i}{\hbar }(x-x_{0}).p}  \tag{4.12}
\end{equation}
$\varphi (x_{0},p)$ is a specific but otherwise arbitrary real function and $%
\Omega \lbrack P(x_{0}\rightarrow x)]$ is the path ordered U(1) holonomy
along the path $P(x_{0}\rightarrow x)$. Notice that $g(x)$ and $h(p)$ are
both necessary because they simultaneously appear in the most general wave
function (4.12). The wave function (4.12) generalizes in a transparent
manner the usual plane wave solutions of application to the trivial
representation of the Heisenberg algebra with $A_{\alpha }(x)=0$ and with
the choices $g(x)=1$ and $h(p)=1$.

Now we point out that the wave-particle duality can be made explicit in
quantum mechanics if we further introduce the equations 
\begin{equation}
\langle p\mid \hat{p}_{\alpha }\mid \psi \rangle =p_{\alpha }\langle p\mid
\psi \rangle =p_{\alpha }\psi (p)  \tag{4.13a}
\end{equation}
\begin{equation}
\langle p\mid \hat{x}_{\alpha }\mid \psi \rangle =\frac{i\hbar }{h^{1/4}(p)}[%
\frac{\partial }{\partial p^{\alpha }}+\frac{i}{\hbar }A_{\alpha }(p)%
]h^{1/4}(p)\psi (p)  \tag{4.13b}
\end{equation}
which are the momentum representation correspondents of equations (4.6). The
vector field $A_{\alpha }(p)$ has a vanishing strength tensor in momentum
space, 
\begin{equation}
F_{\alpha \beta }=\frac{\partial A_{\beta }}{\partial p^{\alpha }}-\frac{%
\partial A_{\alpha }}{\partial p^{\beta }}=0  \tag{4.14}
\end{equation}
and is related to arbitrary local phase transformations of the momentum
eigenvectors 
\begin{equation}
\mid p^{\prime }\rangle =e^{\frac{i}{\hbar }\gamma (p)}\mid p\rangle 
\tag{4.15a}
\end{equation}
when 
\begin{equation}
A_{\alpha }^{\prime }(p)=A_{\alpha }(p)+\frac{\partial \gamma (p)}{\partial p%
}  \tag{4.15b}
\end{equation}
where $\gamma (p)$ is an arbitrary scalar function. Equations (4.15) are the
momentum representation correspondents of equations (4.7).

Now we need further evidence that the complementation of the basic equations
of quantum mechanics we suggested in this section really makes sense. Part
of this evidence comes from the results we derived for the massless particle
in section two. We describe more evidence in the next section,

\section{Dual Fields in Conformal Gravity}

In this section we describe world-line constrained Hamiltonian formalisms
containing tensor and vector fields. These formalisms are constructed
starting from action (3.4b). We will use these formalisms to support our
constructions in section four. These formalisms are based on new local
symmetries in the presence of tensor and vector fields. For the actions we
display here, local $Sp(2,R)$ is a broken symmetry. But this does not cause
any problem in the transition to the physical sector of the theory. This is
because these new local symmetry can also be used to arrive at the correct
number of physical degrees of freedom, and reproduce the local $Sp(2,R)$
symmetry when the fields are removed.

\subsection{Position dependent tensor fields}

Action (3.4b) can be extended to an action in $d+2$ dimensions where the
geometry is described by a position dependent tensor $G_{MN}(X)$. This
action is given by 
\begin{equation*}
S=\int d\tau \{\frac{1}{2}(\dot{X}^{M}P_{M}-X^{M}\dot{P}_{M})-[\frac{1}{2}%
\lambda _{1}G_{MN}(X)P^{M}P^{N}
\end{equation*}
\begin{equation*}
+\lambda _{2}G_{MN}(X)X^{M}P^{N}+\frac{1}{2}\lambda
_{3}G_{MN}(X)X^{M}X^{N}]\}
\end{equation*}
\begin{equation*}
=\int d\tau \{\dot{X}^{M}P_{M}-[\frac{1}{2}\lambda _{1}G_{MN}(X)P^{M}P^{N}
\end{equation*}
\begin{equation}
+\lambda _{2}G_{MN}(X)X^{M}P^{N}+\frac{1}{2}\lambda
_{3}G_{MN}(X)X^{M}X^{N}]\}  \tag{5.1}
\end{equation}
where the Hamiltonian is 
\begin{equation*}
H=\frac{1}{2}\lambda _{1}G_{MN}(X)P^{M}P^{N}+\lambda _{2}G_{MN}(X)X^{M}P^{N}
\end{equation*}
\begin{equation}
+\frac{1}{2}\lambda _{3}G_{MN}(X)X^{M}X^{N}  \tag{5.2}
\end{equation}
The duality transformation (3.13) is not a symmetry of action (5.1) because $%
G_{MN}(X)$ becomes $G_{MN}(P)$ under (3.13).

The equations of motion for the variables $\lambda _{\alpha }$ give the
constraints 
\begin{equation}
\phi _{1}=\frac{1}{2}G_{MN}(X)P^{M}P^{N}\approx 0  \tag{5.3a}
\end{equation}
\begin{equation}
\phi _{2}=G_{MN}(X)X^{M}P^{N}\approx 0  \tag{5.3b}
\end{equation}
\begin{equation}
\phi _{3}=\frac{1}{2}G_{MN}(X)X^{M}X^{N}\approx 0  \tag{5.3c}
\end{equation}
Following Dirac%
\'{}%
s algorithm for constrained systems [20], we must now require the dynamical
stability of constraints (5.3), which is the requirement that 
\begin{equation}
\dot{\phi}_{\alpha }=\{\phi _{\alpha },H\}=\lambda _{\beta }\{\phi _{\alpha
},\phi _{\beta }\}=0  \tag{5.4}
\end{equation}
Conditions (5.4) lead to the Poisson brackets 
\begin{equation}
\{\phi _{1},\phi _{2}\}=-G_{MN}\frac{\partial \phi _{1}}{\partial X_{M}}%
X^{N}-G_{MN}\frac{\partial \phi _{2}}{\partial X_{M}}P^{N}  \tag{5.5a}
\end{equation}
\begin{equation}
\{\phi _{1},\phi _{3}\}=-G_{MN}\frac{\partial \phi _{3}}{\partial X_{M}}P^{N}
\tag{5.5b}
\end{equation}
\begin{equation}
\{\phi _{2},\phi _{3}\}=-G_{MN}\frac{\partial \phi _{3}}{\partial X_{M}}X^{N}
\tag{5.5c}
\end{equation}
The bracket relations (5.5) reproduce the $Sp(2,R)$ gauge algebra (3.9) when 
$G_{MN}=\eta _{MN}$.

We see from (5.5) that we can achieve dynamical stability of constraints
(5.3) if we impose the conditions 
\begin{equation}
G_{MN}(X)\frac{\partial \phi _{\alpha }}{\partial X_{M}}X^{N}\approx 0 
\tag{5.6a}
\end{equation}
\begin{equation}
G_{MN}(X)\frac{\partial \phi _{\alpha }}{\partial X_{M}}P^{N}\approx 0 
\tag{5.6b}
\end{equation}
If we interpret conditions (5.6) as new independent secondary [20]
constrains and require their dynamical stability we get new conditions
involving second order derivatives of the constraints (5.3) with respect to $%
X_{M}$. These new conditions are direct consequences of conditions (5.6). We
therefore retain only conditions (5.6) as the necessary conditions for the
dynamical stability of constraints (5.3). In a transition to flat space $%
G_{MN}=\eta _{MN}$ the conditions (5.6) reproduce the first class
constraints (3.6). When conditions (5.6) hold, constraints (5.3) become
first class constraints.

Constraints (5.3) generate the local transformations 
\begin{equation}
\delta X_{M}=\epsilon _{\alpha }(\tau )\{X_{M},\phi _{\alpha }\}=\epsilon
_{1}G_{MR}P^{R}+\epsilon _{2}G_{MR}X^{R}  \tag{5.7a}
\end{equation}
\begin{equation*}
\delta P_{M}=\epsilon _{\alpha }(\tau )\{P_{M},\phi _{\alpha }\}=-\frac{1}{2}%
\epsilon _{1}\frac{\partial G_{RS}}{\partial X^{M}}P^{R}P^{S}-\epsilon _{2}%
\frac{\partial G_{RS}}{\partial X^{M}}X^{R}P^{S}
\end{equation*}
\begin{equation}
-\epsilon _{2}G_{MR}P^{R}-\frac{1}{2}\epsilon _{3}\frac{\partial G_{RS}}{%
\partial X^{M}}X^{R}X^{S}-\epsilon _{3}G_{MR}X^{R}  \tag{5.7b}
\end{equation}
with arbitrary parameters $\epsilon _{\alpha }(\tau )$. Under
transformations (5.7) action (5.1) transforms as 
\begin{equation*}
\delta S=\int d\tau \lbrack (\epsilon _{1}\lambda _{2}-\epsilon _{2}\lambda
_{1})G_{MN}\frac{\partial \phi _{1}}{\partial X_{M}}X^{N}+(\epsilon
_{2}\lambda _{1}-\epsilon _{1}\lambda _{2})G_{MN}\frac{\partial \phi _{2}}{%
\partial X_{M}}P^{N}
\end{equation*}
\begin{equation*}
+(\epsilon _{3}\lambda _{1}-\epsilon _{1}\lambda _{3})G_{MN}\frac{\partial
\phi _{3}}{\partial X_{M}}P^{N}+(\epsilon _{3}\lambda _{2}-\epsilon
_{2}\lambda _{3})G_{MN}\frac{\partial \phi _{3}}{\partial X_{M}}X^{N}
\end{equation*}
\begin{equation}
+\frac{d}{d\tau }(\epsilon _{\alpha }\phi _{\alpha })+\dot{\epsilon}_{\alpha
}\phi _{\alpha }-\phi _{\alpha }\delta \lambda _{\alpha }]  \tag{5.7c}
\end{equation}
Using the conditions (5.6) we can write the variation (5.7c) as 
\begin{equation}
\delta S\approx \int d\tau \lbrack \frac{d}{d\tau }(\epsilon _{\alpha }\phi
_{\alpha })+\dot{\epsilon}_{\alpha }\phi _{\alpha }-\phi _{\alpha }\delta
\lambda _{\alpha }]  \tag{5.7d}
\end{equation}
If we now choose $\delta \lambda _{\alpha }=\dot{\epsilon}_{\alpha }$ we
obtain 
\begin{equation}
\delta S\approx \int d\tau \frac{d}{d\tau }(\epsilon _{\alpha }\phi _{\alpha
})  \tag{5.7e}
\end{equation}
This confirms that when conditions (5.6) hold action (5.1) has a local
invariance generated by the first class constraints (5.3). The \ conserved
charge corresponding to this local invariance is the quantity 
\begin{equation*}
Q=\epsilon _{\alpha }\phi _{\alpha }=\frac{1}{2}\epsilon
_{1}G_{MN}(X)P^{M}P^{N}+\epsilon _{2}G_{MN}(X)X^{M}P^{N}
\end{equation*}
\begin{equation}
+\frac{1}{2}\epsilon _{3}G_{MN}(X)X^{M}X^{N}  \tag{5.8}
\end{equation}
Since we have here three first class constraints, which is the same number
of first class constraints associated with the local $Sp(2,R)$ symmetry, it
is possible [21] to use the local symmetry generated by the conserved charge
(5.8) to eliminate one space-like degree of freedom and two time-like
degrees of freedom from each of the canonical variables. This leaves us with 
$d-1$ physical canonical pairs. The physical components of the tensor field
are described in terms of the physical components of the position variable.
There will be no ghosts in the quantized theory. As we will see in the next
section, a duality transformation of the type (3.13) changes the local
transformation equations (5.7) and the corresponding conserved charge (5.8)
in the background $G_{MN}(X)$ into the local transformation equations and
the corresponding conserved charge in a background $G_{MN}(P)$. In a
transition to the flat $d+2$ dimensional space-time the transformation
equations (5.7) and the corresponding conserved charge (5.8) reproduce the
local $Sp(2,R)$ transformation equations (3.11) and the corresponding
conserved charge (3.12).

Hamiltonian (5.2) generates the equations of motion 
\begin{equation}
\dot{X}_{M}=\{X_{M},H\}=\lambda _{1}G_{MR}P^{R}+\lambda _{2}G_{MR}X^{R} 
\tag{5.9a}
\end{equation}
\begin{equation*}
\dot{P}_{M}=\{P_{M},H\}=-\frac{1}{2}\lambda _{1}\frac{\partial G_{RS}}{%
\partial X^{M}}P^{R}P^{S}-\lambda _{2}\frac{\partial G_{RS}}{\partial X^{M}}%
X^{R}P^{S}
\end{equation*}
\begin{equation}
-\lambda _{2}G_{MR}P^{R}-\frac{1}{2}\lambda _{3}\frac{\partial G_{RS}}{%
\partial X^{M}}X^{R}X^{S}-\lambda _{3}G_{MR}X^{R}  \tag{5.9b}
\end{equation}
The equations of motion (5.9) reproduce the equations of motion (3.14) when $%
G_{MN}=\eta _{MN}$. As we will see in the next section, the duality
transformation (3.13) changes the equations of motion (5.9) in the
background $G_{MN}(X)$ into the equations of motion in a background $%
G_{MN}(P)$ and vice versa. It is this dual behavior of the local symmetries
and of the equations of motion under the duality transformation (3.13) that
we expect to exist also at the quantum level.

\subsection{Momentum dependent tensor fields}

Now we apply the duality transformation (3.13) to action (5.1) and obtain
the action 
\begin{equation*}
S=\int d\tau \{\frac{1}{2}(\dot{X}^{M}P_{M}-X^{M}\dot{P}_{M})-[\frac{1}{2}%
\lambda _{1}G_{MN}(P)P^{M}P^{N}
\end{equation*}
\begin{equation*}
+\lambda _{2}G_{MN}(P)X^{M}P^{N}+\frac{1}{2}\lambda
_{3}G_{MN}(P)X^{M}X^{N}]\}
\end{equation*}
\begin{equation*}
=\int d\tau \{-X^{M}\dot{P}_{M}-[\frac{1}{2}\lambda _{1}G_{MN}(P)P^{M}P^{N}
\end{equation*}
\begin{equation}
+\lambda _{2}G_{MN}(P)X^{M}P^{N}+\frac{1}{2}\lambda
_{3}G_{MN}(P)X^{M}X^{N}]\}  \tag{5.10}
\end{equation}
where the Hamiltonian is 
\begin{equation*}
H=\frac{1}{2}\lambda _{1}G_{MN}(P)P^{M}P^{N}+\lambda _{2}G_{MN}(P)X^{M}P^{N}
\end{equation*}
\begin{equation}
+\frac{1}{2}\lambda _{3}G_{MN}(P)X^{M}X^{N}  \tag{5.11}
\end{equation}

The equations of motion for the variables $\lambda _{\alpha }(\tau )$ give
the constraints 
\begin{equation}
\phi _{1}=\frac{1}{2}G_{MN}(P)P^{M}P^{N}\approx 0  \tag{5.12a}
\end{equation}
\begin{equation}
\phi _{2}=G_{MN}(P)X^{M}P^{N}\approx 0  \tag{5.12b}
\end{equation}
\begin{equation}
\phi _{3}=\frac{1}{2}G_{MN}(P)X^{M}X^{N}\approx 0  \tag{5.12c}
\end{equation}
Requiring the dynamical stability of constraints (5.12) 
\begin{equation}
\dot{\phi}_{\alpha }=\{\phi _{\alpha },H\}=\lambda _{\beta }\{\phi _{\alpha
},\phi _{\beta }\}=0  \tag{5.13}
\end{equation}
we arrive at the bracket relations 
\begin{equation}
\{\phi _{1},\phi _{2}\}=-G_{MN}\frac{\partial \phi _{1}}{\partial P_{M}}P^{N}
\tag{5.14a}
\end{equation}
\begin{equation}
\{\phi _{1},\phi _{3}\}=-G_{MN}\frac{\partial \phi _{1}}{\partial P_{M}}X^{N}
\tag{5.14b}
\end{equation}
\begin{equation}
\{\phi _{2},\phi _{3}\}=G_{MN}\frac{\partial \phi _{3}}{\partial P_{M}}%
P^{N}-G_{MN}\frac{\partial \phi _{2}}{\partial P_{M}}X^{N}  \tag{5.14c}
\end{equation}
The bracket relations (5.14) reproduce the $Sp(2,R)$ gauge algebra (3.9)
when $G_{MN}=\eta _{MN}$. Therefore, for consistency, we see from equations
(5.14) that to achieve dynamical stability of constraints (5.12) we must
impose the conditions 
\begin{equation}
G_{MN}(P)\frac{\partial \phi _{\alpha }}{\partial P_{M}}X^{N}\approx 0 
\tag{5.15a}
\end{equation}
\begin{equation}
G_{MN}(P)\frac{\partial \phi _{\alpha }}{\partial P_{M}}P^{N}\approx 0 
\tag{5.15b}
\end{equation}
When conditions (5.15) are satisfied, the bracket relations (5.14) weakly
vanish and the constraints (5.12) become first class constraints. Conditions
(5.15) reproduce the $Sp(2,R)$ constraints (3.6) when $G_{MN}=\eta _{MN}$.

Constraints (5.12) generate the local transformations

\begin{equation*}
\delta X_{M}=\epsilon _{\alpha }(\tau )\{X_{M},\phi _{\alpha }\}=\frac{1}{2}%
\epsilon _{1}\frac{\partial G_{RS}}{\partial P^{M}}P^{R}P^{S}+\epsilon
_{1}G_{MR}P^{R}
\end{equation*}
\begin{equation}
+\epsilon _{2}\frac{\partial G_{RS}}{\partial P^{M}}X^{R}P^{S}+\epsilon
_{2}G_{MR}X^{R}+\frac{1}{2}\epsilon _{3}\frac{\partial G_{RS}}{\partial P^{M}%
}X^{R}X^{S}  \tag{5.16a}
\end{equation}
\begin{equation}
\delta P_{M}=\epsilon _{\alpha }(\tau )\{P_{M},\phi _{\alpha }\}=-\epsilon
_{2}G_{MR}P^{R}-\epsilon _{3}G_{MR}X^{R}  \tag{5.16b}
\end{equation}
under which \ 
\begin{equation*}
\delta S=\int d\tau \lbrack (\epsilon _{2}\lambda _{1}-\epsilon _{1}\lambda
_{2})G_{MN}\frac{\partial \phi _{1}}{\partial P_{M}}P^{N}+(\epsilon
_{3}\lambda _{1}-\epsilon _{1}\lambda _{3})G_{MN}\frac{\partial \phi _{2}}{%
\partial P_{M}}P^{N}
\end{equation*}
\begin{equation*}
+(\epsilon _{3}\lambda _{2}-\epsilon _{2}\lambda _{3})G_{MN}\frac{\partial
\phi _{2}}{\partial P_{M}}X^{N}+(\epsilon _{2}\lambda _{3}-\epsilon
_{3}\lambda _{2})G_{MN}\frac{\partial \phi _{3}}{\partial P_{M}}P^{N}
\end{equation*}
\begin{equation}
+\frac{d}{d\tau }(\epsilon _{\alpha }\phi _{\alpha })+\dot{\epsilon}_{\alpha
}\phi _{\alpha }-\phi _{\alpha }\delta \lambda _{\alpha }]  \tag{5.16c}
\end{equation}
When conditions (5.15) hold, we can write the variation of the action as 
\begin{equation}
\delta S\approx \int d\tau \lbrack \frac{d}{d\tau }(\epsilon _{\alpha }\phi
_{\alpha })+\dot{\epsilon}_{\alpha }\phi _{\alpha }-\phi _{\alpha }\delta
\lambda _{\alpha }]  \tag{5.16d}
\end{equation}
If we now choose $\delta \lambda _{\alpha }=\dot{\epsilon}_{\alpha }$ the
variation of the action becomes 
\begin{equation}
\delta S\approx \int d\tau \frac{d}{d\tau }(\epsilon _{\alpha }\phi _{\alpha
})  \tag{5.16e}
\end{equation}
This confirms that, when conditions (5.15) are valid, action (5.10) has a
local invariance generated by the first class constraints (5.12). The
conserved Hamiltonian Noether charge corresponding to this local invariance
is the quantity 
\begin{equation*}
Q=\epsilon _{\alpha }\phi _{\alpha }=\frac{1}{2}\epsilon
_{1}G_{MN}(P)P^{M}P^{N}+\epsilon _{2}G_{MN}(P)X^{M}P^{N}
\end{equation*}
\begin{equation}
+\frac{1}{2}\epsilon _{3}G_{MN}(P)X^{M}X^{N}  \tag{5.17}
\end{equation}
Again it is in principle possible to use the local symmetry generated by the
conserved charge (5.17) to eliminate one space-like degree of freedom and
two time-like degrees of freedom from each of the canonical variables.
Again, we are left with $d-1$ physical canonical pairs and with a tensor
field that depends only on the physical components of the momentum variable.
There will be no ghosts in the quantized theory. If we apply the duality
transformation (3.13a), together with the transformations $\epsilon
_{1}\rightarrow \epsilon _{3}$, \ $\epsilon _{2}\rightarrow -\epsilon _{2}$,
\ $\epsilon _{3}\rightarrow \epsilon _{1}$ to the local transformation
equations (5.16) and to the conserved charge (5.17), they are turned into
the local transformation equations (5.7) and conserved charge (5.8). In a
transition to flat space $G_{MN}=\eta _{MN}$ the transformation equations
(5.16) and the conserved charge (5.17) reproduce the local $Sp(2,R)$
transformations (3.11) and the corresponding conserved Noether charge (3.12).

Hamiltonian (5.11) generates the equations of motion 
\begin{equation*}
\dot{X}_{M}=\{X_{M},H\}=\frac{1}{2}\lambda _{1}\frac{\partial G_{RS}}{%
\partial P^{M}}P^{R}P^{S}+\lambda _{1}G_{MR}P^{R}
\end{equation*}
\begin{equation}
+\lambda _{2}\frac{\partial G_{RS}}{\partial P^{M}}X^{R}P^{S}+\lambda
_{2}G_{MR}X^{R}+\frac{1}{2}\lambda _{3}\frac{\partial G_{RS}}{\partial P^{M}}%
X^{R}X^{S}  \tag{5.18a}
\end{equation}
\begin{equation}
\dot{P}_{M}=\{P_{M},H\}=-\lambda _{2}G_{MR}P^{R}-\lambda _{3}G_{MR}X^{R} 
\tag{5.18b}
\end{equation}
Equations (5.18) reproduce the equations of motion (3.14) when $G_{MN}=\eta
_{MN}$. Under the duality transformation (3.13), the equations of motion
(5.18) in the background $G_{MN}(P)$ are transformed into the equations of
motion (5.9) in the background $G_{MN}(X)$.

\subsection{Position dependent vector fields\ }

Starting from action (3.4b) we can construct the following action [28] in
the presence of a vector field $A_{M}(X)$ 
\begin{equation*}
S=\int d\tau \lbrack \frac{1}{2}(\dot{X}^{M}P_{M}-X^{M}\dot{P}_{M})-(\frac{1%
}{2}\lambda _{1}P^{2}+\lambda _{2}X.P
\end{equation*}
\begin{equation*}
+\frac{1}{2}\lambda _{3}X^{2}+\lambda _{4}X.A+\lambda _{5}P.A+\frac{1}{2}%
\lambda _{6}A^{2})]
\end{equation*}
\begin{equation*}
=\int d\tau \lbrack \dot{X}.P-(\frac{1}{2}\lambda _{1}P^{2}+\lambda _{2}X.P+%
\frac{1}{2}\lambda _{3}X^{2}
\end{equation*}
\begin{equation}
+\lambda _{4}X.A+\lambda _{5}P.A+\frac{1}{2}\lambda _{6}A^{2})]  \tag{5.19}
\end{equation}
where the Hamiltonian is 
\begin{equation*}
H=\frac{1}{2}\lambda _{1}P^{2}+\lambda _{2}X.P+\frac{1}{2}\lambda _{3}X^{2}
\end{equation*}
\begin{equation}
+\lambda _{4}X.A(X)+\lambda _{5}P.A(X)+\frac{1}{2}\lambda _{6}A^{2}(X) 
\tag{5.20}
\end{equation}
The duality transformation (3.13) is not a symmetry of action (3.19) because 
$A_{M}(X)$ becomes $A_{M}(P)$ under (3.13).

The equations of motion for the variables $\lambda _{\varrho }(\tau )$ ($%
\varrho =1,2,...,6$) give the constraints 
\begin{equation}
\phi _{1}=\frac{1}{2}P^{2}\approx 0\text{ \ \ \ \ }\phi _{2}=X.P\approx 0%
\text{ \ \ \ \ }\phi _{3}=\frac{1}{2}X^{2}\approx 0  \tag{5.21a}
\end{equation}
\begin{equation}
\phi _{4}=X.A(X)\approx 0\text{ \ \ \ \ }\phi _{5}=P.A(X)\approx 0\text{ \ \
\ \ }\phi _{6}=\frac{1}{2}A^{2}(X)\approx 0  \tag{5.21b}
\end{equation}
Requiring the dynamical stability of constraints (5.21) 
\begin{equation}
\dot{\phi}_{\varrho }=\{\phi _{\varrho },H\}=\lambda _{\gamma }\{\phi
_{\varrho },\phi _{\gamma }\}=0  \tag{5.22}
\end{equation}
we arrive at the $Sp(2,R)$ brackets (3.9), together with the new brackets 
\begin{equation*}
\{\phi _{1},\phi _{4}\}=-P^{M}\frac{\partial \phi _{4}}{\partial X^{M}}\text{
\ \ \ \ \ \ \ \ \ }\{\phi _{1},\phi _{5}\}=-P^{M}\frac{\partial \phi _{5}}{%
\partial X^{M}}
\end{equation*}
\begin{equation*}
\{\phi _{1},\phi _{6}\}=-P^{M}\frac{\partial \phi _{6}}{\partial X^{M}}\text{
\ \ \ \ \ \ \ \ }\{\phi _{2},\phi _{4}\}=-X^{M}\frac{\partial \phi _{4}}{%
\partial X^{M}}
\end{equation*}
\begin{equation*}
\{\phi _{2},\phi _{5}\}=\phi _{5}-X^{M}\frac{\partial \phi _{5}}{\partial
X^{M}}\text{ \ \ \ \ \ \ }\{\phi _{2},\phi _{6}\}=-X^{M}\frac{\partial \phi
_{6}}{\partial X^{M}}
\end{equation*}
\begin{equation*}
\{\phi _{3},\phi _{4}\}=0\text{ \ \ \ \ \ \ \ \ \ \ \ \ }\{\phi _{3},\phi
_{5}\}=\phi _{4}
\end{equation*}
\begin{equation*}
\{\phi _{3},\phi _{6}\}=0\text{ \ \ \ \ \ \ \ \ \ \ \ \ }\{\phi _{4},\phi
_{5}\}=A^{M}\frac{\partial \phi _{4}}{\partial X^{M}}
\end{equation*}
\begin{equation}
\{\phi _{4},\phi _{6}\}=0\text{ \ \ \ \ \ \ \ \ \ \ \ \ \ \ \ }\{\phi
_{5},\phi _{6}\}=-A^{M}\frac{\partial \phi _{6}}{\partial X^{M}}  \tag{5.23}
\end{equation}
To satisfy the dynamical stability requirement (5.22) we then impose the
conditions 
\begin{equation}
\eta _{MN}\frac{\partial \phi _{\varrho }}{\partial X_{M}}X^{N}\approx 0%
\text{ \ \ \ \ \ \ \ \ \ \ \ }\eta _{MN}\frac{\partial \phi _{\varrho }}{%
\partial X_{M}}P^{N}\approx 0\text{ \ \ \ \ \ \ \ \ \ \ \ }\eta _{MN}\frac{%
\partial \phi _{\varrho }}{\partial X_{M}}A^{N}\approx 0\text{\ \ \ \ } 
\tag{5.24}
\end{equation}
As in the case when only a tensor field is present, conditions (5.24) should
not be interpreted as secondary constraints. They are interpreted here as
restrictions on the motion of the massless particle arising from the
presence of constraints (5.21). When these conditions are valid the
constraints (5.21) become first class constraints.

Action (5.19) is invariant under global Lorentz $SO(d,2)$ transformation
with generator $L_{MN}=X_{M}P_{N}-X_{N}P_{M}$ 
\begin{equation}
\delta X_{M}=\frac{1}{2}\omega _{RS}\{L_{RS},X_{M}\}=\omega _{MR}X_{R} 
\tag{5.25a}
\end{equation}
\begin{equation}
\delta P_{M}=\frac{1}{2}\omega _{RS}\{L_{RS},P_{M}\}=\omega _{MR}P_{R} 
\tag{5.25b}
\end{equation}
\begin{equation}
\delta A_{M}=\frac{\partial A_{M}}{\partial X_{R}}\delta X_{R}  \tag{5.25c}
\end{equation}
\begin{equation}
\delta \lambda _{\varrho }=0,\text{ \ \ \ \ \ }\varrho =1,2,...,6 
\tag{5.25d}
\end{equation}
under which $\delta S=0$. It can be checked that $L_{MN}$ has weakly
vanishing Poisson brackets with constraints (5.21), being therefore gauge
invariant in the presence of the vector field $A_{M}(X)$.

Constraints (5.21) generate the local transformations 
\begin{equation}
\delta X_{M}=\epsilon _{\varrho }(\tau )\{X_{M},\phi _{\varrho }\}=\epsilon
_{1}P_{M}+\epsilon _{2}X_{M}+\epsilon _{5}A_{M}  \tag{5.26a}
\end{equation}
\begin{equation*}
\delta P_{M}=\epsilon _{\varrho }(\tau )\{P_{M},\phi _{\varrho }\}=-\epsilon
_{2}P_{M}-\epsilon _{3}X_{M}-\epsilon _{4}A_{M}
\end{equation*}
\begin{equation}
-\epsilon _{4}X^{N}\frac{\partial A_{N}}{\partial X^{M}}-\epsilon _{5}P^{N}%
\frac{\partial A_{N}}{\partial X^{M}}-\epsilon _{6}A^{N}\frac{\partial A_{N}%
}{\partial X^{M}}  \tag{5.26b}
\end{equation}
Using conditions (5.24), the variation of the action becomes 
\begin{equation*}
\delta S\approx \int d\tau \lbrack (\epsilon _{2}\lambda _{1}-\epsilon
_{1}\lambda _{2})2\phi _{1}+(\epsilon _{3}\lambda _{1}-\epsilon _{1}\lambda
_{3})\phi _{2}
\end{equation*}
\begin{equation*}
+(\epsilon _{3}\lambda _{2}-\epsilon _{2}\lambda _{3})2\phi _{3}+(\epsilon
_{3}\lambda _{5}-\epsilon _{5}\lambda _{3})\phi _{4}
\end{equation*}
\begin{equation}
+(\epsilon _{2}\lambda _{5}-\epsilon _{5}\lambda _{2})\phi _{5}+\frac{d}{%
d\tau }(\epsilon _{\varrho }\phi _{\varrho })+\dot{\epsilon}_{\varrho }\phi
_{\varrho }-\phi _{\varrho }\delta \lambda _{\varrho }]  \tag{5.26c}
\end{equation}
Using now the constraint equations (5.21), we find that this variation
reduces to 
\begin{equation}
\delta S\approx \int d\tau \lbrack \frac{d}{d\tau }(\epsilon _{\varrho }\phi
_{\varrho })+\dot{\epsilon}_{\varrho }\phi _{\varrho }-\phi _{\varrho
}\delta \lambda _{\varrho }]  \tag{5.26d}
\end{equation}
If we then finally choose $\delta \lambda _{\varrho }=\dot{\epsilon}%
_{\varrho }$ we are left with 
\begin{equation}
\delta S\approx \int d\tau \frac{d}{d\tau }(\epsilon _{\varrho }\phi
_{\varrho })  \tag{5.26e}
\end{equation}
Equation (5.26e) shows that the conserved Hamiltonian Noether charge in a
flat $d+2$ dimensional space-time with the presence of a background vector
field $A_{M}(X)$ is the quantity 
\begin{equation*}
Q=\epsilon _{\varrho }\phi _{\varrho }=\frac{1}{2}\epsilon
_{1}P^{2}+\epsilon _{2}X.P+\frac{1}{2}\epsilon _{3}X^{2}
\end{equation*}
\begin{equation}
+\epsilon _{4}X.A(X)+\epsilon _{5}P.A(X)+\frac{1}{2}\epsilon _{6}A^{2}(X) 
\tag{5.27}
\end{equation}
We can use the local symmetry generated by the conserved charge (5.27) to
eliminate one space-like and two time-like degrees of freedom from each of
the canonical variables and from the vector field. We are then left with a
physical phase space with $d-1$ canonical pairs over which is defined a
vector field with $d-1$ physical components. The vector field depends only
on the physical components of the position variable. Again, there will be no
ghosts in the quantized theory. The transformation equations (5.26) and the
corresponding conserved charge (5.27) reproduce the local $Sp(2,R)$
transformation equations (3.11) and the corresponding conserved charge
(3.12) when $A_{M}(X)=0$.

The Hamiltonian equations of motion in the presence of the vector field $%
A_{M}(X)$ are 
\begin{equation}
\dot{X}_{M}=\{X_{M},H\}=\lambda _{1}P_{M}+\lambda _{2}X_{M}+\lambda _{5}A_{M}
\tag{5.28a}
\end{equation}
\begin{equation*}
\dot{P}_{M}=\{P_{M},H\}=-\lambda _{2}P_{M}-\lambda _{3}X_{M}-\lambda
_{4}A_{M}
\end{equation*}
\begin{equation}
-\lambda _{4}X^{N}\frac{\partial A_{N}}{\partial X^{M}}-\lambda _{5}P^{N}%
\frac{\partial A_{N}}{\partial X^{M}}-\lambda _{6}A^{N}\frac{\partial A_{N}}{%
\partial X^{M}}  \tag{5.28b}
\end{equation}
Clearly these equations of motion reproduce the equations of motion (3.14)
when $A_{M}(X)=0$.

\subsection{Momentum dependent vector fields}

Now we perform in action (3.19) the duality transformation 
\begin{equation}
X_{M}(\tau )\rightarrow P_{M}(\tau )\text{ \ \ \ \ \ }P_{M}(\tau
)\rightarrow -X_{M}(\tau )  \tag{5.29a}
\end{equation}
\begin{equation}
\lambda _{1}\rightarrow \lambda _{3}\text{ \ \ \ \ }\lambda _{2}\rightarrow
-\lambda _{2}\text{ \ \ \ \ }\lambda _{3}\rightarrow \lambda _{1} 
\tag{5.29b}
\end{equation}
\begin{equation}
\lambda _{4}\rightarrow \lambda _{5}\text{ \ \ \ \ }\lambda _{5}\rightarrow
-\lambda _{4}\text{ \ \ \ \ }\lambda _{6}\rightarrow \lambda _{6} 
\tag{5.29c}
\end{equation}
and obtain the following action with a background vector field $A_{M}(P)$ 
\begin{equation*}
S=\int d\tau \lbrack \frac{1}{2}(\dot{X}^{M}P_{M}-X^{M}\dot{P}_{M})-(\frac{1%
}{2}\lambda _{1}P^{2}+\lambda _{2}X.P
\end{equation*}
\begin{equation*}
+\frac{1}{2}\lambda _{3}X^{2}+\lambda _{4}X.A+\lambda _{5}P.A+\frac{1}{2}%
\lambda _{6}A^{2})]
\end{equation*}
\begin{equation*}
=\int d\tau \lbrack \dot{X}^{M}P_{M}-(\frac{1}{2}\lambda _{1}P^{2}+\lambda
_{2}X.P+\frac{1}{2}\lambda _{3}X^{2}
\end{equation*}
\begin{equation}
+\lambda _{4}X.A+\lambda _{5}P.A+\frac{1}{2}\lambda _{6}A^{2})]  \tag{5.30}
\end{equation}
\ where the Hamiltonian is 
\begin{equation*}
H=\frac{1}{2}\lambda _{1}P^{2}+\lambda _{2}X.P+\frac{1}{2}\lambda _{3}X^{2}
\end{equation*}
\begin{equation}
+\lambda _{4}X.A(P)+\lambda _{5}P.A(P)+\frac{1}{2}\lambda _{6}A^{2}(P) 
\tag{5.31}
\end{equation}
The equations of motion for the variables $\lambda _{\varrho }(\tau )$ give
the constraints 
\begin{equation}
\phi _{1}=\frac{1}{2}P^{2}\approx 0\text{ \ \ \ \ }\phi _{2}=X.P\approx 0%
\text{ \ \ \ \ }\phi _{3}=\frac{1}{2}X^{2}\approx 0  \tag{5.32a}
\end{equation}
\begin{equation}
\phi _{4}=X.A(P)\approx 0\text{ \ \ \ \ }\phi _{5}=P.A(P)\approx 0\text{ \ \
\ \ }\phi _{6}=\frac{1}{2}A^{2}(P)\approx 0  \tag{5.32b}
\end{equation}
Requiring the dynamical stability condition (5.22) for constraints (5.32) we
arrive at the new bracket relations 
\begin{equation*}
\{\phi _{1},\phi _{4}\}=-\phi _{5}\text{ \ \ \ \ \ }\{\phi _{1},\phi _{5}\}=0
\end{equation*}
\begin{equation*}
\{\phi _{1},\phi _{6}\}=0\text{ \ \ \ \ \ }\{\phi _{2},\phi _{4}\}=-\phi
_{4}+P^{M}\frac{\partial \phi _{4}}{\partial P^{M}}
\end{equation*}
\begin{equation*}
\{\phi _{2},\phi _{5}\}=P^{M}\frac{\partial \phi _{5}}{\partial P^{M}}\text{
\ \ \ \ \ }\{\phi _{2},\phi _{6}\}=P^{M}\frac{\partial \phi _{6}}{\partial
P^{M}}
\end{equation*}
\begin{equation*}
\{\phi _{3},\phi _{4}\}=X^{M}\frac{\partial \phi _{4}}{\partial P^{M}}\text{
\ \ \ \ \ }\{\phi _{3},\phi _{5}\}=X^{M}\frac{\partial \phi _{5}}{\partial
P^{M}}
\end{equation*}
\begin{equation*}
\{\phi _{3},\phi _{6}\}=X^{M}\frac{\partial \phi _{6}}{\partial P^{M}}\text{
\ \ \ \ \ }\{\phi _{4},\phi _{5}\}=A^{M}\frac{\partial \phi _{5}}{\partial
P^{M}}
\end{equation*}
\begin{equation}
\{\phi _{4},\phi _{6}\}=A^{M}\frac{\partial \phi _{6}}{\partial P^{M}}\text{
\ \ \ \ \ }\{\phi _{5},\phi _{6}\}=0  \tag{5.33}
\end{equation}
We see from (5.33) that to obtain dynamical stability for constraints (5.32)
we must impose the conditions 
\begin{equation}
\eta _{MN}\frac{\partial \phi _{\varrho }}{\partial P_{M}}X^{N}\approx 0%
\text{ \ \ \ \ \ }\eta _{MN}\frac{\partial \phi _{\varrho }}{\partial P_{M}}%
P^{N}\approx 0\text{ \ \ \ \ \ }\eta _{MN}\frac{\partial \phi _{\varrho }}{%
\partial P_{M}}A^{N}\approx 0  \tag{5.34}
\end{equation}
When conditions (5.34) hold, constraints (5.32) become first class
constraints.

Action (5.30) has a global Lorentz $SO(d,2)$ invariance with generator $%
L_{MN}=X_{M}P_{N}-X_{N}P_{M}$ 
\begin{equation}
\delta X_{M}=\frac{1}{2}\omega _{RS}\{L_{RS},X_{M}\}=\omega _{MR}X_{R} 
\tag{5.35a}
\end{equation}
\begin{equation}
\delta P_{M}=\frac{1}{2}\omega _{RS}\{L_{RS},P_{M}\}=\omega _{MR}P_{R} 
\tag{5.35b}
\end{equation}
\begin{equation}
\delta A_{M}=\frac{\partial A_{M}}{\partial P_{N}}\delta P_{N}  \tag{5.35c}
\end{equation}
\begin{equation}
\delta \lambda _{\varrho }=0\text{ \ \ \ \ \ }\varrho =1,2,.....,6 
\tag{5.35d}
\end{equation}
under which $\delta S=0$. It can be verified that $L_{MN}$ has weakly
vanishing Poisson brackets with constraints (5.32) being therefore also
gauge invariant in the presence of the vector field $A_{M}(P)$.

Constraints (5.32) generate the local transformations 
\begin{equation*}
\delta X_{M}=\epsilon _{\varrho }(\tau )\{X_{M},\phi _{\varrho }\}=\epsilon
_{1}P_{M}+\epsilon _{2}X_{M}+\epsilon _{4}X^{S}\frac{\partial A_{S}}{%
\partial P^{M}}
\end{equation*}
\begin{equation}
+\epsilon _{5}A_{M}+\epsilon _{5}P^{S}\frac{\partial A_{S}}{\partial P^{M}}%
+\epsilon _{6}A^{S}\frac{\partial A_{S}}{\partial P^{M}}  \tag{5.36a}
\end{equation}
\begin{equation}
\delta P_{M}=\epsilon _{\varrho }(\tau )\{P_{M},\phi _{\varrho }\}=-\epsilon
_{2}P_{M}-\epsilon _{3}X_{M}-\epsilon _{4}A_{M}  \tag{5.36b}
\end{equation}
\begin{equation}
\delta A_{M}=\frac{\partial A_{M}}{\partial P_{N}}\delta P_{N}  \tag{5.36c}
\end{equation}
under which, after using the conditions (5.34), action (5.30) varies as 
\begin{equation*}
\delta S\approx \int d\tau \lbrack (\epsilon _{2}\lambda _{1}-\epsilon
_{1}\lambda _{2})2\phi _{1}+(\epsilon _{3}\lambda _{1}-\epsilon _{1}\lambda
_{3})\phi _{2}
\end{equation*}
\begin{equation*}
+(\epsilon _{3}\lambda _{2}-\epsilon _{2}\lambda _{3})2\phi _{3}+(\epsilon
_{4}\lambda _{2}-\epsilon _{2}\lambda _{4})\phi _{4}
\end{equation*}
\begin{equation}
+(\epsilon _{4}\lambda _{1}-\epsilon _{1}\lambda _{4})\phi _{5}+\frac{d}{%
d\tau }(\epsilon _{\varrho }\phi _{\varrho })+\dot{\epsilon}_{\varrho }\phi
_{\varrho }-\phi _{\varrho }\delta \lambda _{\varrho }]  \tag{5.36d}
\end{equation}
Using the constraint equations (5.32), the above variation becomes 
\begin{equation}
\delta S\approx \int d\tau \lbrack \frac{d}{d\tau }(\epsilon _{\varrho }\phi
_{\varrho })+\dot{\epsilon}_{\varrho }\phi _{\varrho }-\phi _{\varrho
}\delta \lambda _{\varrho }]  \tag{5.36e}
\end{equation}
If we now choose $\delta \lambda _{\varrho }=\dot{\epsilon}_{\varrho }$ we
obtain 
\begin{equation}
\delta S\approx \int d\tau \frac{d}{d\tau }(\epsilon _{\varrho }\phi
_{\varrho })  \tag{5.36d}
\end{equation}
The conserved charge is then the quantity 
\begin{equation*}
Q=\epsilon _{\varrho }\phi _{\varrho }=\frac{1}{2}\epsilon
_{1}P^{2}+\epsilon _{2}X.P+\frac{1}{2}\epsilon _{3}X^{2}
\end{equation*}
\begin{equation}
+\epsilon _{4}X.A(P)+\epsilon _{5}P.A(P)+\frac{1}{2}\epsilon _{6}A^{2}(P) 
\tag{5.37}
\end{equation}
It is possible to use the local symmetry generated by the conserved charge
(5.37) to eliminate the unphysical degrees of freedom. At the end we are
left with a physical phase space with $d-1$ canonical pairs over which is
defined a vector field with $d-1$ physical components. The vector field
depends only on the physical components of the momentum variable. If we
perform the duality transformation (5.29a), complemented with the
transformations $\epsilon _{1}\rightarrow \epsilon _{3}$, \ $\epsilon
_{2}\rightarrow -\epsilon _{2}$, \ $\epsilon _{3}\rightarrow \epsilon _{1}$,
\ $\epsilon _{4}\rightarrow \epsilon _{5}$, \ $\epsilon _{5}\rightarrow
-\epsilon _{4}$, \ $\epsilon _{6}\rightarrow \epsilon _{6}$, in the local
transformation equations (5.36) and in the conserved charge (5.37), they are
turned into the local transformation equations (5.26) and conserved charge
(5.27). Therefore, there are duality transformations that relate the local
symmetry of the action in the presence of the vector field $A_{M}(P)$ to the
local symmetry of the action in the presence of the vector field $A_{M}(X)$
and vice versa. The transformation equations (5.36) and the corresponding
conserved charge (5.37) reproduce the local $Sp(2,R)$ transformation
equations (3.11) and the corresponding conserved charge (3.12) when $%
A_{M}(P)=0$.

Hamiltonian (5.31) generates the equations of motion 
\begin{equation*}
\dot{X}_{M}=\{X_{M},H\}=\lambda _{1}P_{M}+\lambda _{2}X_{M}+\lambda _{5}A_{M}
\end{equation*}
\begin{equation}
+\lambda _{4}X^{N}\frac{\partial A_{N}}{\partial P^{M}}+\lambda _{5}P^{N}%
\frac{\partial A_{N}}{\partial P^{M}}+\lambda _{6}A^{N}\frac{\partial A_{N}}{%
\partial P^{M}}  \tag{5.38a}
\end{equation}
\begin{equation}
\dot{P}_{M}=\{P_{M},H\}=-\lambda _{2}P_{M}-\lambda _{3}X_{M}-\lambda
_{4}A_{M}  \tag{5.38b}
\end{equation}
Equations of motion (5.38) reproduce the equations of motion (3.14) when $%
A_{M}(P)=0$. If we perform the duality transformation (5.29) in the
equations of motion (5.38), they are turned into the equations of motion
(5.28). The duality transformation (5.29) therefore relates the dynamical
evolution of the massless particle in the vector field $A_{M}(P)$ to its
dynamical evolution in the vector field $A_{M}(X)$ and vice versa.

\subsection{Tensor and vector fields}

As a final way to give evidence that the complementation of quantum
mechanics we suggested in section four makes sense, we now construct two
dual actions with tensor and vector fields. The first of these actions
describes the case when $G_{MN}=G_{MN}(X)$ and $A_{M}=A_{M}(X)$. This action
is 
\begin{equation*}
S=\int d\tau \{\dot{X}^{M}P_{M}-[\frac{1}{2}\lambda
_{1}G_{MN}(X)P^{M}P^{N}+\lambda _{2}G_{MN}(X)X^{M}P^{N}
\end{equation*}
\begin{equation*}
+\frac{1}{2}\lambda _{3}G_{MN}(X)X^{M}X^{N}+\lambda _{4}G_{MN}(X)X^{M}A^{N}
\end{equation*}
\begin{equation}
+\lambda _{5}G_{MN}(X)P^{M}A^{N}+\frac{1}{2}\lambda
_{6}G_{MN}(X)A^{M}A^{N}]\}  \tag{5.39}
\end{equation}
where the Hamiltonian is 
\begin{equation*}
H=\frac{1}{2}\lambda _{1}G_{MN}(X)P^{M}P^{N}+\lambda _{2}G_{MN}(X)X^{M}P^{N}
\end{equation*}
\begin{equation*}
+\frac{1}{2}\lambda _{3}G_{MN}(X)X^{M}X^{N}+\lambda _{4}G_{MN}(X)X^{M}A^{N}
\end{equation*}
\begin{equation}
+\lambda _{5}G_{MN}(X)P^{M}A^{N}+\frac{1}{2}\lambda _{6}G_{MN}(X)A^{M}A^{N} 
\tag{5.40}
\end{equation}
The equations of motion for the variables $\lambda _{\varrho }$ give the
constraints 
\begin{equation}
\phi _{1}=\frac{1}{2}G_{MN}(X)P^{M}P^{N}\approx 0\text{ \ \ \ \ \ }\phi
_{2}=G_{MN}(X)X^{M}P^{N}\approx 0  \tag{5.41a}
\end{equation}
\begin{equation}
\phi _{3}=\frac{1}{2}G_{MN}(X)X^{M}X^{N}\text{ }\approx 0\text{ \ \ \ \ \ }%
\phi _{4}=G_{MN}(X)X^{M}A^{N}(X)\approx 0  \tag{5.41b}
\end{equation}
\begin{equation}
\phi _{5}=G_{MN}(X)P^{M}A^{N}(X)\approx 0  \tag{5.41c}
\end{equation}
\begin{equation}
\phi _{6}=\frac{1}{2}G_{MN}(X)A^{M}(X)A^{N}(X)\approx 0  \tag{5.41d}
\end{equation}
Requiring dynamical stability of constraints (5.41) we arrive at the
brackets 
\begin{equation*}
\{\phi _{1},\phi _{2}\}=-G_{MN}\frac{\partial \phi _{1}}{\partial X_{M}}%
X^{N}-G_{MN}\frac{\partial \phi _{2}}{\partial X_{M}}P^{N}
\end{equation*}
\begin{equation*}
\{\phi _{1},\phi _{3}\}=-G_{MN}\frac{\partial \phi _{3}}{\partial X_{M}}P^{N}
\end{equation*}
\begin{equation*}
\{\phi _{2},\phi _{3}\}=-G_{MN}\frac{\partial \phi _{3}}{\partial X_{M}}X^{N}
\end{equation*}
\begin{equation*}
\{\phi _{1},\phi _{4}\}=-G_{MN}\frac{\partial \phi _{4}}{\partial X_{M}}P^{N}
\end{equation*}
\begin{equation*}
\{\phi _{1},\phi _{5}\}=G_{MN}\frac{\partial \phi _{1}}{\partial X_{M}}%
A^{N}-G_{MN}\frac{\partial \phi _{5}}{\partial X_{M}}P^{N}
\end{equation*}
\begin{equation*}
\{\phi _{1},\phi _{6}\}=-G_{MN}\frac{\partial \phi _{6}}{\partial X_{M}}P^{N}
\end{equation*}
\begin{equation*}
\{\phi _{2},\phi _{4}\}=-G_{MN}\frac{\partial \phi _{4}}{\partial X_{M}}X^{N}
\end{equation*}
\begin{equation*}
\{\phi _{2},\phi _{5}\}=G_{MN}\frac{\partial \phi _{5}}{\partial X_{M}}%
A^{N}-G_{MN}\frac{\partial \phi _{5}}{\partial X_{M}}X^{N}
\end{equation*}
\begin{equation*}
\{\phi _{2},\phi _{6}\}=-G_{MN}\frac{\partial \phi _{3}}{\partial X_{M}}X^{N}
\end{equation*}
\begin{equation*}
\{\phi _{3},\phi _{5}\}=G_{MN}\frac{\partial \phi _{3}}{\partial X_{M}}A^{N}
\end{equation*}
\begin{equation*}
\{\phi _{4},\phi _{5}\}=G_{MN}\frac{\partial \phi _{4}}{\partial X_{M}}A^{N}
\end{equation*}
\begin{equation}
\{\phi _{5},\phi _{6}\}=-G_{MN}\frac{\partial \phi _{6}}{\partial X_{M}}A^{N}
\tag{5.42}
\end{equation}
To achieve dynamical stability of constraints (5.51) we then impose the
conditions 
\begin{equation}
G_{MN}(X)\frac{\partial \phi _{\varrho }}{\partial X_{M}}X^{N}\approx 0 
\tag{5.43a}
\end{equation}
\begin{equation}
G_{MN}(X)\frac{\partial \phi _{\varrho }}{\partial X_{M}}P^{N}\approx 0 
\tag{5.43b}
\end{equation}
\begin{equation}
G_{MN}(X)\frac{\partial \phi _{\varrho }}{\partial X_{M}}A^{N}\approx 0 
\tag{5.43c}
\end{equation}
If we make a transition to flat space $G_{MN}=\eta _{MN}$ equations (5.42)
reproduce the bracket relations (3.9) and (5.23) for the constraints (5.21)
in the presence of the vector field $A_{M}(X)$ only. If we remove the vector
field by setting $A_{M}(X)=0$ the brackets (5.42) turn into the bracket
relations (5.5) for the constraints (5.3) in the presence of the tensor
field $G_{MN}(X)$ only. If both the tensor and vector fields are removed
then equations (5.42) turn into the $Sp(2,R)$ bracket relations (3.9) for
the first class constraints (3.6).

Constraints (5.41) generate the local transformations 
\begin{equation*}
\delta X_{M}=\epsilon _{\varrho }(\tau )\{X_{M},\phi _{\varrho }\}=\epsilon
_{1}G_{MR}P^{R}+\epsilon _{2}G_{MR}X^{R}
\end{equation*}
\begin{equation}
+\epsilon _{5}G_{MR}A^{R}  \tag{5.44a}
\end{equation}
\begin{equation*}
\delta P_{M}=\epsilon _{\varrho }(\tau )\{P_{M},\phi _{\varrho }\}=-\frac{1}{%
2}\epsilon _{1}\frac{\partial G_{RS}}{\partial X^{M}}P^{R}P^{S}
\end{equation*}
\begin{equation*}
-\epsilon _{2}\frac{\partial G_{RS}}{\partial X^{M}}X^{R}P^{S}-\epsilon
_{2}G_{MR}P^{R}-\frac{1}{2}\epsilon _{3}\frac{\partial G_{RS}}{\partial X^{M}%
}X^{R}X^{S}
\end{equation*}
\begin{equation*}
-\epsilon _{3}G_{MR}X^{R}-\epsilon _{4}\frac{\partial G_{RS}}{\partial X^{M}}%
X^{R}A^{S}-\epsilon _{4}G_{MR}A^{R}
\end{equation*}
\begin{equation*}
-\epsilon _{4}G_{RS}X^{R}\frac{\partial A^{S}}{\partial X^{M}}-\epsilon _{5}%
\frac{\partial G_{RS}}{\partial X^{M}}P^{R}A^{S}-\epsilon _{5}G_{RS}P^{R}%
\frac{\partial A^{S}}{\partial X^{M}}
\end{equation*}
\begin{equation}
-\frac{1}{2}\epsilon _{6}\frac{\partial G_{RS}}{\partial X^{M}}%
A^{R}A^{S}-\epsilon _{6}G_{RS}A^{R}\frac{\partial A^{S}}{\partial X^{M}} 
\tag{5.44b}
\end{equation}
\begin{equation}
\delta A_{M}=\frac{\partial A_{M}}{\partial X^{R}}\delta X^{R}  \tag{5.44c}
\end{equation}
\begin{equation}
\delta G_{MN}=\frac{\partial G_{MN}}{\partial X^{R}}\delta X^{R}  \tag{5.44d}
\end{equation}
under which, after using conditions (5.43), action (5.39) varies as 
\begin{equation}
\delta S\approx \int d\tau \lbrack \frac{d}{d\tau }(\epsilon _{\varrho }\phi
_{\varrho })+\dot{\epsilon}_{\varrho }\phi _{\varrho }-\phi _{\varrho
}\delta \lambda _{\varrho }]  \tag{5.44e}
\end{equation}
By choosing $\delta \lambda _{\varrho }=\dot{\epsilon}_{\varrho }$ the
variation (5.44e) becomes 
\begin{equation}
\delta S\approx \int d\tau \frac{d}{d\tau }(\epsilon _{\alpha }\phi _{\alpha
})  \tag{5.44f}
\end{equation}
This shows that the quantity 
\begin{equation*}
Q=\epsilon _{\varrho }\phi _{\varrho }=\frac{1}{2}\epsilon
_{1}G_{MN}(X)P^{M}P^{N}+\epsilon _{2}G_{MN}(X)X^{M}P^{N}
\end{equation*}
\begin{equation*}
+\frac{1}{2}\epsilon _{3}G_{MN}(X)X^{M}X^{N}+\epsilon
_{4}G_{MN}(X)X^{M}A^{N}(X)
\end{equation*}
\begin{equation}
+\epsilon _{5}G_{MN}(X)P^{M}A^{N}(X)+\frac{1}{2}\epsilon
_{6}G_{MN}(X)A^{M}(X)A^{N}(X)  \tag{5.45}
\end{equation}
is the conserved Hamiltonian Noether charge in the presence of the tensor
field $G_{MN}(X)$ and the vector field $A_{M}(X)$. It is possible to use the
local symmetry generated by the conserved \ charge (5.45) to reach a
physical phase space with $d-1$ canonical pairs over which a vector field
with $d-1$ physical components is defined. The physical components of the
tensor and vector fields depend only on the physical components of the
position variable. There will be no ghosts in the quantized theory.

Hamiltonian (5.40) generates the equations of motion 
\begin{equation*}
\dot{X}_{M}=\{X_{M},H\}=\lambda _{1}G_{MR}P^{R}+\lambda _{2}G_{MR}X^{R}
\end{equation*}
\begin{equation}
+\lambda _{5}G_{MR}A^{R}  \tag{5.46a}
\end{equation}
\begin{equation*}
\dot{P}_{M}=\{P_{M},H\}=-\frac{1}{2}\lambda _{1}\frac{\partial G_{RS}}{%
\partial X^{M}}P^{R}P^{S}-\lambda _{2}\frac{\partial G_{RS}}{\partial X^{M}}%
X^{R}P^{S}
\end{equation*}
\begin{equation*}
-\lambda _{2}G_{MR}P^{R}-\frac{1}{2}\lambda _{3}\frac{\partial G_{RS}}{%
\partial X^{M}}X^{R}X^{S}-\lambda _{3}G_{MR}X^{R}
\end{equation*}
\begin{equation*}
-\lambda _{4}\frac{\partial G_{RS}}{\partial X^{M}}X^{R}A^{S}-\lambda
_{4}G_{MR}A^{R}-\lambda _{4}G_{RS}X^{R}\frac{\partial A^{S}}{\partial X^{M}}
\end{equation*}
\begin{equation*}
-\lambda _{5}\frac{\partial G_{RS}}{\partial X^{M}}P^{R}A^{S}-\lambda
_{5}G_{RS}P^{R}\frac{\partial A^{S}}{\partial X^{M}}-\frac{1}{2}\lambda _{6}%
\frac{\partial G_{RS}}{\partial X^{M}}A^{R}A^{S}
\end{equation*}
\begin{equation}
-\lambda _{6}G_{RS}A^{R}\frac{\partial A^{S}}{\partial X^{M}}  \tag{5.46b}
\end{equation}
If we make a transition to flat space-time, the equations of motion (5.46)
reduce to the equations of motion (5.28) in the presence of the vector field 
$A_{M}(X)$ only. If we set $A_{M}(X)=0$ the equations of motion (5.46)
reduce to the equations of motion (5.9) in the presence of the tensor field $%
G_{MN}(X)$ only. If we set $G_{MN}=\eta _{MN}$ and $A_{M}(X)=0$ the
equations of motion (5.46) reduce to the conformal gravity equations of
motion (3.14).

Performing the duality transformation (5.29) in action (5.39), we turn to
the case when $G_{MN}=G_{MN}(P)$ and $A_{M}=A_{M}(P)$. The action is 
\begin{equation*}
S=\int d\tau \{-X^{M}\dot{P}_{M}-[\frac{1}{2}\lambda
_{1}G_{MN}(P)P^{M}P^{N}+\lambda _{2}G_{MN}(P)X^{M}P^{N}
\end{equation*}
\begin{equation*}
+\frac{1}{2}\lambda _{3}G_{MN}(P)X^{M}X^{N}+\lambda _{4}G_{MN}(P)X^{M}A^{N}
\end{equation*}
\begin{equation}
+\lambda _{5}G_{MN}(P)P^{M}A^{N}+\frac{1}{2}\lambda
_{6}G_{MN}(P)A^{M}A^{N}]\}  \tag{5.47}
\end{equation}
where the Hamiltonian is 
\begin{equation*}
H=\frac{1}{2}\lambda _{1}G_{MN}(P)P^{M}P^{N}+\lambda _{2}G_{MN}(P)X^{M}P^{N}
\end{equation*}
\begin{equation*}
+\frac{1}{2}\lambda _{3}G_{MN}(P)X^{M}X^{N}+\lambda _{4}G_{MN}(P)X^{M}A^{N}
\end{equation*}
\begin{equation}
+\lambda _{5}G_{MN}(P)P^{M}A^{N}+\frac{1}{2}\lambda _{6}G_{MN}(P)A^{M}A^{N} 
\tag{5.48}
\end{equation}
The equations of motion for the $\lambda _{\varrho }(\tau )$ give the
constraints 
\begin{equation}
\phi _{1}=\frac{1}{2}G_{MN}(P)P^{M}P^{N}\approx 0\text{ \ \ \ \ \ }\phi
_{2}=G_{MN}(P)X^{M}P^{N}\approx 0  \tag{5.49a}
\end{equation}
\begin{equation}
\phi _{3}=\frac{1}{2}G_{MN}(P)X^{M}X^{N}\approx 0\text{ \ \ \ \ \ }\phi
_{4}=G_{MN}(P)X^{M}A^{N}(P)\approx 0  \tag{5.49b}
\end{equation}
\begin{equation}
\phi _{5}=G_{MN}(P)P^{M}A^{N}(P)\approx 0  \tag{5.49c}
\end{equation}
\begin{equation}
\phi _{6}=\frac{1}{2}G_{MN}(P)A^{M}(P)A^{N}(P)\approx 0  \tag{5.49d}
\end{equation}
It can be verified that constraints (5.49) become dynamically stable first
class constraints when the conditions 
\begin{equation}
G_{MN}(P)\frac{\partial \phi _{\varrho }}{\partial P_{M}}X^{N}\approx 0 
\tag{5.50a}
\end{equation}
\begin{equation}
G_{MN}(P)\frac{\partial \phi _{\varrho }}{\partial P_{M}}P^{N}\approx 0 
\tag{5.50b}
\end{equation}
\begin{equation}
G_{MN}(P)\frac{\partial \phi _{\varrho }}{\partial P_{M}}A^{N}\approx 0 
\tag{5.50c}
\end{equation}
hold. Constraints (5.49) generate local transformations that can be obtained
from the transformation equations (5.44) by using the duality transformation
(5.29) with the $\epsilon _{\varrho }$ in the place of the $\lambda
_{\varrho }$. In this case the conserved Hamiltonian Noether charge is the
quantity 
\begin{equation*}
Q=\epsilon _{\varrho }\phi _{\varrho }=\frac{1}{2}\epsilon
_{1}G_{MN}(P)P^{M}P^{N}+\epsilon _{2}G_{MN}(P)X^{M}P^{N}
\end{equation*}
\begin{equation*}
+\frac{1}{2}\epsilon _{3}G_{MN}(P)X^{M}X^{N}+\epsilon
_{4}G_{MN}(P)X^{M}A^{N}(P)
\end{equation*}
\begin{equation}
+\epsilon _{5}G_{MN}(P)P^{M}A^{N}(P)+\frac{1}{2}\epsilon
_{6}G_{MN}(P)A^{M}(P)A^{N}(P)  \tag{5.51}
\end{equation}
We can again in principle use the local symmetry generated by the conserved
charge (5.51) to reach the physical sector of the theory, with $d-1$
canonical pairs and $d-1$ components of the vector field. The physical
components of the tensor and vector fields will depend only on the physical
components of the momentum variable. If we perform the duality
transformation (5.29) with the $\epsilon _{\varrho }$ in the place of the $%
\lambda _{\varrho }$ we find that the conserved charge (5.51) is turned into
the conserved charge (5.45).

Hamiltonian (5.48) generates the equations of motion 
\begin{equation*}
\dot{X}_{M}=\{X_{M},H\}=\frac{1}{2}\lambda _{1}\frac{\partial G_{RS}}{%
\partial P^{M}}P^{R}P^{S}+\lambda _{1}G_{MR}P^{R}
\end{equation*}
\begin{equation*}
+\lambda _{2}\frac{\partial G_{RS}}{\partial P^{M}}X^{R}P^{S}+\lambda
_{2}G_{MR}X^{R}+\frac{1}{2}\lambda _{3}\frac{\partial G_{RS}}{\partial P^{M}}%
X^{R}X^{S}
\end{equation*}
\begin{equation*}
+\lambda _{4}\frac{\partial G_{RS}}{\partial P^{M}}X^{R}A^{S}+\lambda
_{4}G_{RS}X^{R}\frac{\partial A^{S}}{\partial P^{M}}+\lambda _{5}\frac{%
\partial G_{RS}}{\partial P^{M}}P^{R}A^{S}
\end{equation*}
\begin{equation*}
+\lambda _{5}G_{MR}A^{R}+\lambda _{5}G_{RS}P^{R}\frac{\partial A^{S}}{%
\partial P^{M}}+\frac{1}{2}\lambda _{6}\frac{\partial G_{RS}}{\partial P^{M}}%
A^{R}A^{S}
\end{equation*}
\begin{equation}
+\lambda _{6}G_{RS}A^{R}\frac{\partial A^{S}}{\partial P^{M}}  \tag{5.52a}
\end{equation}
\begin{equation*}
\dot{P}_{M}=\{P_{M},H\}=-\lambda _{2}G_{MR}P^{R}-\lambda _{3}G_{MR}X^{R}
\end{equation*}
\begin{equation}
-\lambda _{4}G_{MR}A^{R}  \tag{5.52b}
\end{equation}
If we perform a transition to flat space-time, the equations of motion
(5.52) reduce to the equations of motion (5.38) in the presence of the
vector field $A_{M}(P)$ only. If we set $A_{M}(P)=0$ the equations of motion
(5.52) reduce to the equations of motion (5.18) in the presence of the
tensor field $G_{MN}(P)$ only. If we set $G_{MN}=\eta _{MN}$ and $A_{M}(P)=0$
the equations of motion (5.52) reduce to the conformal gravity equations of
motion (3.14). The equations of motion (5.52) in the presence of the tensor
field $G_{MN}(P)$ and vector field $A_{M}(P)$ are transformed by the duality
transformation (5.29) into the equations of motion (5.46) in the presence of
the tensor field $G_{MN}(X)$ and vector field $A_{M}(X).$

These results indicate that there are two dual descriptions of the local
symmetry and of the dynamical evolution of the massless relativistic
particle in the presence of tensor and vector fields. These two dual
descriptions are related by duality transformations that interchange
position and momentum. These two dual descriptions reduce to conformal
gravity on the world-line when the tensor and vector fields are removed. It
is the existence of these two dual descriptions that suggests the
complementation of quantum mechanics we described in section four.

\section{ Concluding \ remarks}

In this paper we suggested a complementation of the basic equations of
quantum mechanics in the presence of gravity. The new equations we suggested
here are based on a duality transformation that interchanges position and
momentum at the classical level. At the quantum level, this duality
transformation leaves invariant the definition of the fundamental commutator 
$[X_{M},P_{N}]=ih\eta _{MN}$. It is then reasonable to expect that this
duality transformation at the quantum level will change position dependent
tensor and vector fields into momentum dependent tensor and vector fields
without any modification of the fundamental aspects of the quantum dynamics.
In this paper we made an attempt to describe a particular classical limit of
this dual behavior of the dynamics in the presence of tensor and vector
fields, starting from the conformal gravity action in the world-line
formalism, and using constrained Hamiltonian methods.

\end{document}